\documentclass[floatfix,%
 reprint,
 amsmath,amssymb,
 aps,
]{revtex4-2}
\usepackage{lipsum}
\usepackage{tabularx} 
\usepackage[dvipsnames]{xcolor}
\usepackage{graphicx}% Include figure files
\usepackage{dcolumn}% Align table columns on decimal point
\usepackage{bm}% bold math
\usepackage{hyperref}% add hypertext capabilities
\usepackage[mathlines]{lineno}% Enable numbering of text and display math
\usepackage{mathtools}
\usepackage{soul}
\usepackage{float}% Improved float placement
\begin{document}

\preprint{APS/123-QED}

\title{Micromagnetic-atomistic hybrid modeling of defect-induced magnetization dynamics}

\author {Nastaran Salehi$^1$}

\author{Olle Eriksson$^{1,2}$}

\author {Johan Hellsvik$^3$}
 
\author{Manuel Pereiro$^1$}

\affiliation{$^1$Department of Physics and Astronomy, Uppsala University, Box 516, SE-751 20 Uppsala, Sweden \\
$^2$WISE - Wallenberg Initiative Materials Science for Sustainability, Department of Physics and Astronomy, Uppsala University, SE-751 20 Uppsala, Sweden\\ 
$^3$PDC Center for High Performance Computing, KTH Royal Institute of Technology, SE-100 44 Stockholm, Sweden}

\date{\today}

\begin{abstract}

This study presents a multiscale approach for investigating magnetization dynamics in multiscale, hybrid micromagnetic-atomistic simulations. We considered the dynamics of spin waves, domain walls, as well as three-dimensional (3D) skyrmions, in the presence of defects. Two primary defect configurations are examined: (i) a double-slit structure, which enables the study of domain wall and spin wave interference, and (ii) a tetrahedron shaped cluster of atoms with tunable anisotropy, which provides insights into how localized anisotropic perturbations influence domain wall pinning and skyrmion stability. The magnonic double-slit experiment demonstrates interference patterns analogous to electronic wave phenomena, offering potential applications in wave-based computing. Additionally, the results reveal the impact of the local anisotropy that leads to distinct transformations, including domain wall deformations, tubular and spherical structures, skyrmion annihilation, and breathing mode. The findings underscore the critical role of defect-induced anisotropic interactions in controlling domain wall motion, skyrmion topology, and spin wave propagation.

\end{abstract}

\maketitle

%\tableofcontents

\section{\label{sec:level1} Introduction}

The field of spintronics has gained significant interest due to its potential to revolutionize information storage and processing technologies by using the spin degree of freedom of electrons \cite{Zutic2004, Fert2008, Parkin2015}. In particular, domain walls and skyrmions have emerged as promising candidates for next-generation spintronic devices due to their unique topological properties, stability, and energy-efficient manipulation \cite{Dupe2016, Beach2008, Sampaio2013, Nagaosa2013, Wiesendanger2016}, although skyrmion-based devices have not yet been commercialized. Domain walls and skyrmions provide robust mechanisms for data transport and logic operations \cite{Zhang2015, Fert2017, Kruger2012}. However, the behavior of these magnetic structures in the presence of defects and localized anisotropy remains a crucial area of research, as material imperfections and engineered anisotropic regions significantly influence their dynamics \cite{Hillebrands2006}. To explore these effects, it is essential to move beyond ideal models and account for the role of defects and imperfections present in real materials.

While ideal magnetic systems offer promising device functionalities, real materials inevitably contain defects that can fundamentally alter their performance. Recent experimental studies have shown that introduced defects can be used to control magnetic textures. For example, atomic-scale vacancies and dislocations influence skyrmion stability~\cite{buttner2018}, artificial pinning centers guide domain wall motion~\cite{bauer2020}, and tailored anisotropy landscapes enable new switching mechanisms~\cite{gross2017}. These studies reveal that defects are not merely imperfections but powerful tools for controlling magnetization dynamics. However, a systematic multiscale understanding of how atomic-scale defects influence mesoscale magnetic textures remains challenging, particularly for three-dimensional systems.

A key challenge in micromagnetic simulations is the need for multiscale modeling to accurately capture both atomic-scale interactions and large-scale magnetic texture evolution \cite{DeLucia2016}. In this work, we develop and apply the $\mu$-ASD module of the UppASD software \cite{Skubic2008}, a computational framework that integrates atomistic spin dynamics and micromagnetic simulations, enabling a comprehensive investigation of defect-induced effects. This investigation is focused on iron-iridium (Fe-Ir) thin films. The interatomic coupling interactions for these thin films are adopted from \cite{Heinze2011}. This approach allows us to explore localized defect interactions while maintaining computational efficiency in larger-scale magnetic domains.

This study also represents an implementation of a fully three-dimensional (3D) multiscale simulation method, thus increasing its functionality beyond the previous two-dimensional (2D) implementation and analysis \cite{Mendez2020}. The present research extends the paradigm to explore 3D magnetization dynamics. This advancement facilitates a better comprehension of skyrmion behavior, domain wall interactions, and spin wave propagation within realistic, multilayered, and confined magnetic systems. As a result, and as demonstrated here, this progress opens up new prospects for multiscale modeling in 3D geometries.

We focus primarily on two type of defects, namely a double-slit and an atomic cluster with tetrahedral geometry. A double-slit structure placed in the middle of the material serves as an ideal platform to investigate domain wall scattering and spin wave interference, similarly as in the case of the electronic double-slit experiment in optics and quantum mechanics. The double-slit experiment demonstrates the wave-particle duality of quantum particles, where a coherent wavefront passing through two slits produces an interference pattern on the other side. Similarly, in this work, we explore the magnonic counterpart of the double-slit experiment, where spin waves, excited by a microwave field, propagate through a %structured 
magnetic medium containing a double-slit and generate an interference pattern characteristic of wave-like behavior. Additionally, we examine the dynamics of a domain wall crossing the double slit, where an external magnetic field is applied to drive the domain wall along the longitudinal direction. As the domain wall interacts with the spatial constraints of the slits, its trajectory and shape are altered, similar to how particles and waves in the quantum double-slit experiment experience diffraction and interference effects. A dual analysis of spin wave interference and domain wall dynamics in a double-slit structure provides valuable insight into how structured defects affect magnetization dynamics. The results contribute to the field of magnonics, where controlled interference of spin waves can be utilized for, e.g, information processing, magnonic transistors, and wave-based computing architectures \cite{Chumak2015, Kruglyak2010}. This study not only establishes an intriguing link between quantum mechanical wave phenomena and magnonic wave dynamics but also provides a practical framework for designing spin wave-based logic and memory devices using engineered defect structures.

The second type of defect considered in this study, as aforementioned, is an atomic cluster with tetrahedral geometry, with tunable anisotropy. Here, a localized tetrahedral anisotropic region is introduced to examine its effect on domain wall pinning and skyrmion dynamics. By systematically tuning the anisotropy strength and orientation (in-plane vs. out-of-plane orientation), we analyze how these modifications influence skyrmion and domain wall deformation \cite{Crum2015}.
Our results show that increasing the anisotropy in a small region (here described as a tetrahedron cluster) induces distinct magnetic texture responses, including domain wall elongation, skyrmion annihilation, and skyrmion breathing. Furthermore, the magnonic double-slit experiment reveals wave interference patterns that demonstrate the feasibility of controlling spin wave propagation using engineered defects. These findings contribute to a broader understanding of defect engineering in spintronic materials.

The paper is organized as follows: In Section II, we describe the method, detailing the Landau-Lifshitz-Gilbert (LLG)-based computational framework and the multiscale integration of atomistic and micromagnetic regions. Section III presents the results and analysis, discussing the impact of localized anisotropic perturbations on domain walls, skyrmions, and spin waves. Finally, Section IV provides concluding remarks, summarizing the implications of this study for future spintronic applications and defect engineering strategies. The supplementary information includes videos that illustrate the temporal evolution of the systems analyzed in this study.

\section{Methods}

This study continues the development of the $\mu$-ASD module of UppASD, first introduced in Ref. \cite{Mendez2020}. This is a specialized simulation package for spin dynamics that connects the atomistic and micromagnetic scale. The $\mu$-ASD module is based on the LLG equation, which governs the temporal evolution of magnetic moments under several interactions, as discussed below. The multiscale nature of the method facilitates the integration of atomistic spin dynamics and micromagnetic simulations, enabling the study of systems where localized atomic-scale phenomena, such as defects, coexist with larger micromagnetic structures like domain walls and skyrmions. The relevant parameters for both scales are derived from the Heisenberg Hamiltonian.

For atomistic interactions, the LLG equation in the form of Landau-Lifshitz (LL), describing the dynamics of an atomic spin is given by:

\begin{equation}
\frac{\mathrm{d} \mathbf{m}_i}{\mathrm{d}t} = - \frac{\gamma}{1+\alpha^2}(\mathbf{m}_i \times \mathbf{H}_i^\mathrm{eff} + \frac{\alpha}{m}\mathbf{m}_i \times (\mathbf{m}_i \times \mathbf{H}_i^\mathrm{eff}))
\label{eq:1}
\end{equation}
where $\gamma$ is the gyromagnetic ratio, $\mathbf{m_i}$ is the magnetic moment of atom $i$, $m$ is modulus of magnetic moment, and $\alpha$ is the Gilbert damping parameter. The effective field  incorporates multiple contributions:

\begin{align}
\mathbf{H}^\mathrm{eff}       &= \mathbf{H}^{\mathrm{exc}} + \mathbf{H}^{\mathrm{DM}} + \mathbf{H}^{\mathrm{ani}} + \mathbf{H}^{\mathrm{Z}} \label{eq:2} \\ % (2)
\mathbf{H}_i^{\mathrm{exc}}   &= -\sum_{j \neq i} J_{ij} \mathbf{m}_j \label{eq:3} \\                               % (3)
\mathbf{H}_i^{\mathrm{DM}}     &= -\sum_{j \neq i} \mathbf{D}_{ij} \times \mathbf{m}_j \label{eq:4} \\              % (4)
\mathbf{H}_i^{\mathrm{ani}}   &= -2K_a \mathbf{m}_i \label{eq:5} \\                                                % (5)
\mathbf{H}^{\mathrm{Z}}       &= -g \mu_B \mathbf{B}_\mathrm{ext} \label{eq:6}                                       % (6)
\end{align}
where $\mu_B$ is the Bohr magneton, $g$ is the g-factor, and the different terms in Eq.~\ref{eq:2} represents the Heisenberg exchange, the Dzyaloshinskii-Moriya (DM), the anisotropy and external magnetic fields, respectively. These fields are governed by the interaction couplings ${J}_{ij}$ (Heisenberg exchange), $\mathbf{D}_{ij}$ (DMI), anisotropy constant $K_a$, and magnetic field $\mathbf{B}_\mathrm{ext}$.

The micromagnetic dynamics is described by the micromagnetic LL differential equation as \cite{POLUEKTOV2018219}:

\begin{equation}
 \frac{\partial \mathbf{m}}{\partial t} =- \frac{\gamma}{1+\alpha^2}( \mathbf{m} \times \mathbf{H}^\mathrm{eff} + \alpha \frac{\mathbf{m}}{m} \times (\mathbf{m} \times {\mathbf{H}^\mathrm{eff}}))
 \label{eq:7}
\end{equation}
where $\mathbf{m}$ is the magnetization, a continuous function of space and time. The effective field consists of multiple contributions:

\begin{align}
\mathbf{H}^\mathrm{eff}   &= \mathbf{H}^\mathrm{exc} + \mathbf{H}^\mathrm{DM} + \mathbf{H}^\mathrm{ani} + \mathbf{H}^\mathrm{Z} \label{eq:8} \\
\mathbf{H}^\mathrm{exc}   &= -\bar{A}_{e} : \nabla \nabla \mathbf{m} \label{eq:9} \\
\mathbf{H}^\mathrm{DM}     &= -\nabla \cdot \left(\bar{D}_e \times \mathbf{m}\right) \label{eq:10} \\
\mathbf{H}^\mathrm{ani}   &= -2K_a \mathbf{m} \label{eq:11} \\
\mathbf{H}^\mathrm{Z}     &= -g\mu_B \mathbf{B}_\mathrm{ext} \label{eq:12}
\end{align}
with Eqs.~\ref{eq:9}-\ref{eq:12} representing the exchange field,  the DMI contribution,  the anisotropy field, and  the Zeeman field due to an external magnetic field, respectively. The micromagnetic parameters include the exchange stiffness $\bar{A}_{e}$, DM interaction $\bar{D}_e$, anisotropy constant $K_a$, and external field represented by $\mathbf{B}_\mathrm{ext}$. The matrix $\bar{A}_e$ is diagonal, while $\bar{D}_e$ is a antisymmetric matrix with zeros on its diagonal elements. In Eq.~\ref{eq:9}, two tensors are multiplied by the double dot product which results in following expression \cite{Gibbs1901}: 
\begin{equation}
    A:B=\sum_{\alpha=1}^3 \sum_{\beta=1}^3 A_{\alpha \beta} B_{ \beta \alpha}.
    \label{eq:13}
\end{equation}
In making a connection to  Eq.~\ref{eq:9}, we note that in Eq.~\ref{eq:13} $A$ represents $\bar{A}_e$, and $B$ represents the operator $\nabla \nabla$.

Atomistic and micromagnetic models differ in their representation of magnetization. Micromagnetics treats magnetization as a continuous field, while atomistic models consider discrete spin moments. To ensure consistency, a mapping between atomistic and micromagnetic exchange parameters is established. The exchange stiffness and DMI tensor  in the micromagnetic framework are related to their atomistic counterparts as:

\begin{align}
\bar{A}_e           &= \frac{1}{2} \sum_{j \neq i} J_{ij} \mathbf{r}_{ij}\mathbf{r}_{ij} \label{eq:14} \\
\bar{D}_e     &= \sum_{j \neq i} \mathbf{r}_{ij} \mathbf{D}_{ij} \label{eq:15} \\
\mathbf{r}_{ij} &= \mathbf{r}_{j} - \mathbf{r}_{i} \label{eq:16}
\end{align}
where $\mathbf{r}_{ij}$ is the vector connecting atom $i$ with atom $j$. In materials with spatially varying exchange coefficients, micromagnetic parameters are computed separately for different atomic environments and averaged accordingly to ensure consistency across scales. This multiscale approach enables accurate modeling of magnetic materials with heterogeneous exchange interactions and DMI anisotropies.

In multiscale simulations one integrates the atomistic and continuum domains, where each region follows its respective governing equations. To facilitate connection between these regions, a specialized technique is employed to solve the relevant differential equations. A smooth transition at the interface is ensured by introducing padding atoms and padding nodes within their respective domains \cite{POLUEKTOV2018219}. Unlike conventional magnetization formulations, these padding elements acquire their magnetic moments through interpolation from their respective domains, thereby maintaining smoothness at the atomistic-continuum interface. Additionally, within the atomistic region, coarse-graining techniques may be applied near the continuum interface, leading to the formation of coarse-graining and damping bands to enhance computational efficiency and stability \cite{POLUEKTOV2018219}.

In this study, the micromagnetic system under investigation is an iron-iridium (Fe-Ir) composite. This system is characterized by nearest neighbor atomic interactions, with an exchange interaction value of 0.419 mRy, a Dzyaloshinskii-Moriya interaction (DMI) of 0.132 mRy, and a uniaxial anisotropy of 0.059 mRy. These values have been taken from experimental measurements as it was reported in Ref. \cite{Heinze2011}. We note that the nearest neighbor approximation adopted here for the atomistic region is not a limitation of the formalism, presented above, but rather a practical choice for the presently investigated system. 
In all simulations presented here, we consider a larger micromagnetic framework with an atomistic region in the middle, enabling a high-resolution analysis of atomic-scale phenomena. The corresponding micromagnetic parameters when calculated from Eqs.~\ref{eq:14}-\ref{eq:15}, will have $A_{e}$ with a value of 0.627 mRy/\AA, ${D}_e$ with a value of 0.099 mRy/\AA$^2$, and the anisotropy field, which is the same as for the atomistic region, as it represents a local field. These parameters were kept the same for all simulations presented here, except for smaller regions in the atomic domain, where we in some cases considered local impurities with different magnetic anisotropy.

A primary focus of this study is the introduction of a localized defect within the atomistic region. By embedding such a small-scale defect into the significantly larger micromagnetic system, the aim is to investigate how the defect influences the overall magnetic behavior and dynamics of the system. This study particularly emphasizes understanding the interaction of the defect with specific magnetic configurations, namely domain walls and skyrmions. These configurations are chosen due to their fundamental importance in spintronics and their sensitivity to local perturbations. In addition, this allows to study the dynamics of magnetic objects with trivial and non-trivial topology. The introduction of defects in the middle of the system provides an ideal platform to examine how localized structural or magnetic variations can alter domain wall motion, skyrmion lifetime, and overall magnetic properties.

In this study, the calculation of the skyrmion topological charge in 3D micromagnetic simulations has been successfully implemented within the computational framework. This enhancement extends conventional 2D skyrmion charge calculations to fully 3D systems, allowing for a more general characterization of skyrmion tubes, and other 3D topological textures \cite{Borisov2024, Fert2017, Nagaosa2013, Drissi2021}. In the micromagnetic continuum limit, the topological skyrmion number in 2D is defined as:

\begin{equation}
Q^S = \frac{1}{4\pi} \int \mathbf{m} \cdot \left( \frac{\partial \mathbf{m}}{\partial x} \times \frac{\partial \mathbf{m}}{\partial y} \right) d^2r
\label{eq:17}
\end{equation}
where $\mathbf{m}$ is the normalized magnetization vector. This equation is typically valid in 2D micromagnetic systems, where the skyrmion charge describes the number of times the magnetization field wraps around the unit sphere \cite{Zhang2015, Sampaio2013}. However, for fully 3D skyrmion tubes, additional generalizations are required, incorporating spatial variations along the third dimension \cite{Muller2016, Rybakov2013}. The topological skyrmion number in 3D is defined as \cite{Zhang2015, Stavrou2019, Lee2009}:

\begin{equation}
Q^V = \frac{1}{8\pi} \int \mathbf{m} \cdot \left( \frac{\partial \mathbf{m}}{\partial x} \times \frac{\partial \mathbf{m}}{\partial y} \right) d^3r.
\label{eq:18}
\end{equation}
By generating a 3D skyrmion, the computed topological skyrmion number is confirmed to be $Q^V = 1$, consistent with the expected topological invariant for an isolated skyrmion. This implementation enables further investigations into 3D skyrmion dynamics and expanding the micromagnetic simulations in topological magnetism.

This work aims to uncover the fundamental mechanisms by which defects in hybrid micromagnetic-atomistic systems impact the stability, dynamics, and structure of magnetic configurations. 
To achieve this, the defect is treated at an atomistic level, providing a high-resolution view of its interactions and influence. The surrounding regions, which are free of defects, are modeled using a micromagnetic approach. This hybrid modeling approach enables a multiscale analysis, combining the benefits of atomistic precision in the defect region with the computational efficiency of micromagnetic simulations for the larger system.

The study comprises two distinct series of simulations, each designed to probe the response of the system under different initial magnetic configurations. In the first series, a domain wall is introduced within the micromagnetic structure. The interaction of the domain wall with the atomistic defect region is analyzed to elucidate the impact of the defect on domain wall dynamics, structure, and stability. In the second series, the system is initialized with a 3D skyrmion, a topologically protected magnetic structure. Here, the focus is on understanding how the defect alters the skyrmion properties, such as its size, lifetime, and dynamics. 

This dual-approach methodology provides a comprehensive understanding of defect-induced effects in micromagnetic systems. It not only sheds light on the fundamental mechanisms governing defect interactions in hybrid micromagnetic-atomistic systems but also has potential implications for designing advanced magnetic materials and devices where precise defect engineering plays a crucial role.

\section{Results}

\subsection{A magnonic double-slit system}

In this study, a double-slit described in the following is introduced in the middle of the system in the atomistic region, with the purpose of investigating the interference pattern of magnons.

\subsubsection{Spin Wave interference}

The primary objective of this analysis is to examine the interference pattern of spin waves as they propagate through the slits, drawing an analogy to the classical electronic double-slit experiment in optics and quantum mechanics. In conventional electronic systems, the double-slit experiment demonstrates the wave-particle duality of electrons, revealing an interference pattern indicative of their wave-like behavior. Similarly, in this study, we aim to explore the magnonic equivalent of the double-slit experiment, where spin waves of collective spin excitations form interference pattern as they pass through the two slits. 

\begin{figure}[h]
    \centering    
\includegraphics[width=0.43\textwidth, height=0.4\textwidth]{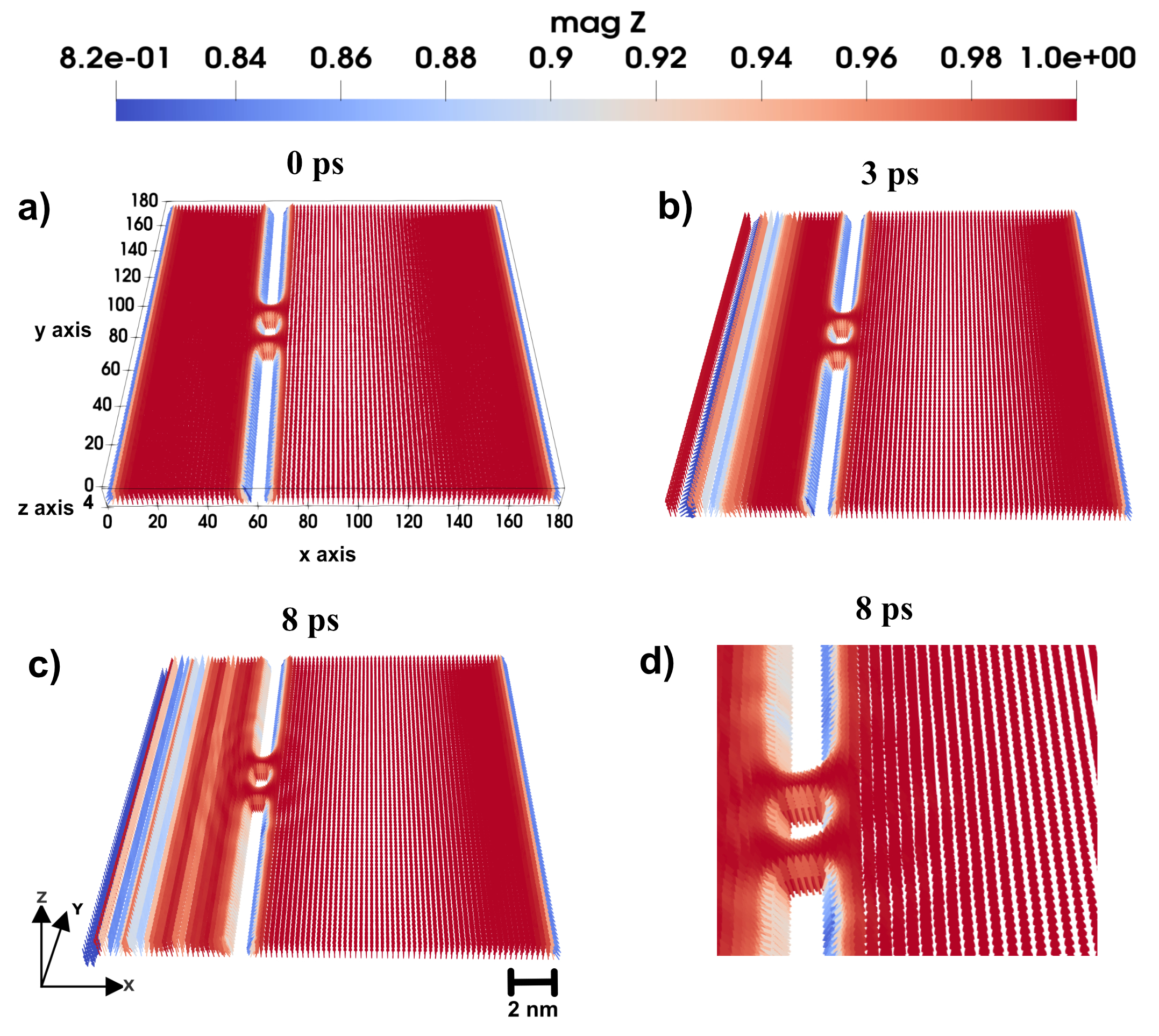}
    \caption{Spin wave propagation through a 3D double-slit. In a) a 3D perspective of the system setup is illustrated together with the magnetic configuration in 0~ps time. In b) and c), we have 3 and 8~ps time frames of the system dynamics, respectively. A magnification of panel c) around the slits is displayed in panel d) to illustrate the interference pattern. The color code is showing the $z$ component of normalized magnetization (red +1 and blue 0.82).}
    \label{fig:1}
\end{figure}

By analyzing the simulated interference patterns, this work provides fundamental insights into the wave nature of magnons and establishes parallels between quantum mechanical wave interference and magnonic wave dynamics, which develops new spin-based applications. The double-slit structure (shown in Fig.~\ref{fig:1}) serves as a controlled environment to study scattering effects and possible pinning mechanisms that arise due to spatially defined perturbations. In this part of the study, we investigate the propagation of spin waves through a double-slit configuration by applying a local oscillating microwave field to excite spin waves of the micromagnetic region. We include this term in LLG equation as \cite{Taniguchi2016}:

\begin{equation}
\mathbf{H}_{{MW}}(t) = \mathbf{H}_0 \cos(w t + \phi)
\label{eq:19}
\end{equation}
where $w=2\pi f$ is angular frequency and $f$ stands for the microwave frequency. The time variable is defined by $t$. $H_0$ is the amplitude of the microwave field, and $\phi$ is the phase of the microwave field.

For the Fe-Ir system, the magnon excitations are in an energy range up to 100 meV \cite{Zakeri2017}, which corresponds to a valid frequency range up to 24 THz. In our simulation, the excitation frequency is set to be 1 THz, ensuring a realistic representation of magnon dynamics within the system. The amplitude is set to be 1500 T, the pulse duration is defined as $300$ ps, and the phase of the microwave field is chosen to be 1.5. The pulse is switched on and off at the proper time without using a Gaussian envelope. While the amplitude value used in simulations exceeds experimentally realistic magnitudes, it was intentionally set high to rapidly perturb spin configurations and thereby reduce the computational time. Additionally, this choice still preserves the underlying physics of the system. These parameters are optimized to effectively excite and sustain magnonic oscillations.

The field induced in Eq.~\ref{eq:19} generates coherent spin waves, which travel through the micromagnetic system, interacting with the double-slit region in the middle of the system (see Fig.~\ref{fig:1}). The micromagnetic system under investigation has dimensions of $180 \times 180 \times 4$~\AA$^3$. The atomistic region has the size of $10 \times 180 \times 4$~\AA$^3$. The rectangular regions, in the absence of  magnetic moments, have the size of $5 \times 75 \times 4$~\AA$^3$~ (the upper and lower ones in Fig.~\ref{fig:1}) and $5 \times 10 \times 4$~\AA$^3$~ (the middle one in Fig.~\ref{fig:1}). The slits should be narrow enough to act as coherent sources and spaced optimally to produce a clear interference pattern, therefore the width of each slit is set to 10~\AA. Slits are positioned at  57 \AA~ in $x$-direction and the distance from center to center of the slits is 20~\AA. 

Figure~\ref{fig:1} illustrates the spin wave propagation through the double slit. The dynamics of the system is presented in the Supplementary Material, Video 1 \cite{Supplementary}.

The interference of spin waves traveling through the two slits is visible from Fig.~\ref{fig:1} d) and seems to share common features to optical waves or the quantum mechanical double-slit experiment, but differs in the underlying physics and governing principles. In the quantum case, as specified by the Schrödinger equation, interference arises from the wavefunction of particles, whereas in spin waves, as results of the LLG equation, interference results from collective magnon excitations in a magnetically ordered medium that according to the results in Fig.~\ref{fig:1} have wave-like properties. 

\begin{figure}[h]
    \centering    
\includegraphics[width=0.45\textwidth, height=0.21\textwidth]{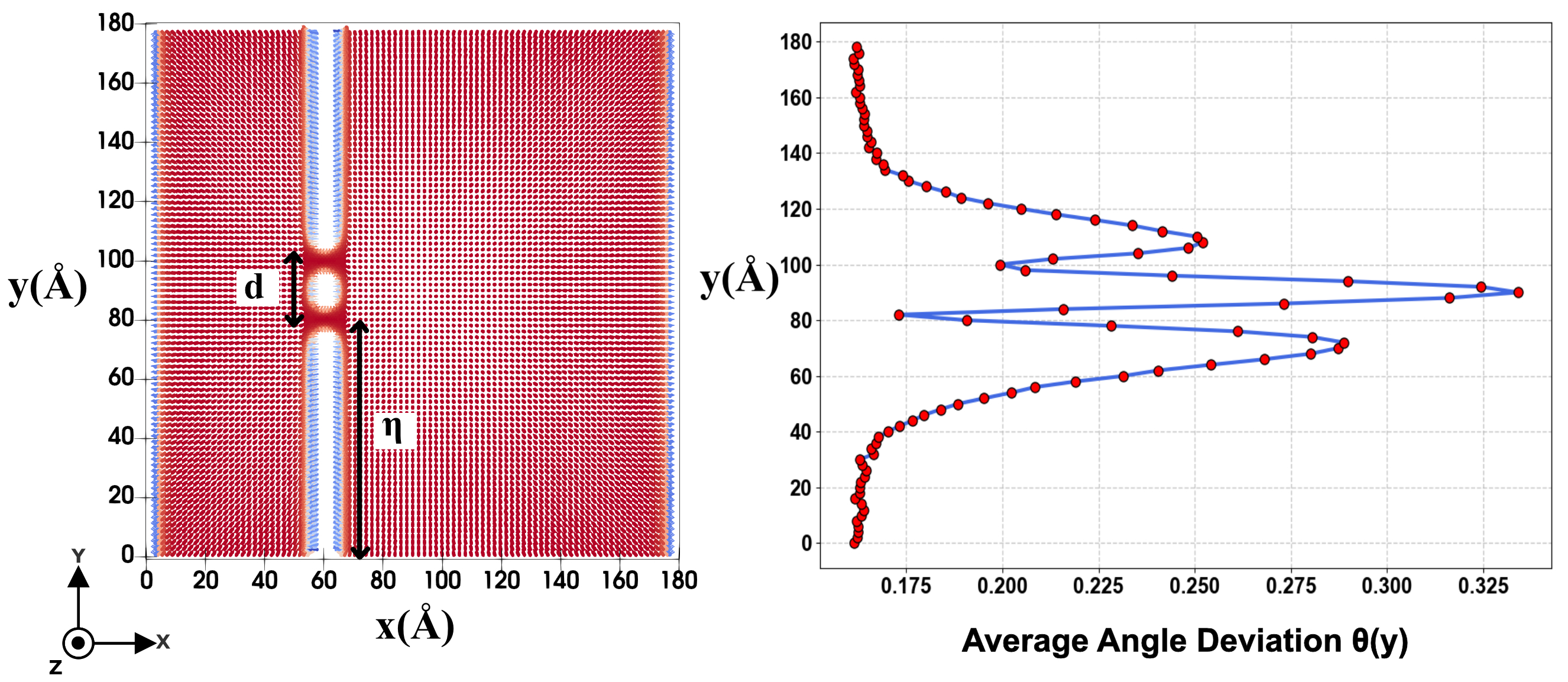}
    \caption{Micromagnetic interference pattern (right figure) induced by average angle deviation of the moment over time from $z$-direction, in degrees, as a function of the length along the $y$-axis (geometry shown in the left plot) at a chosen value of $x=82$~\AA~, $z=2$~\AA,  $d=20$~\AA~ (the distance from the center to center of slits), and $\eta$=80~\AA~ (the distance from $y=0$~\AA~ to the center of the lower slit).}
    
    \label{fig:2}
\end{figure}

Based on the micromagnetic simulation results of the double slit system, interference pattern is observed as illustrated in Fig.~\ref{fig:2}. The right-hand plot displays the interference pattern characterized by the average angular deviation of the magnetic moment over time from the $z$-direction, measured in degrees, as a function of the position along the $y$-axis. This analysis is conducted at a fixed coordinate of $x = 82$~\AA~ and $z = 2$~\AA, where the intensity and peak of the interference pattern reach their maximum.

In order to make a proper comparison between the micromagnetic simulation results and theoretical predictions, we develop an analytical model for interference pattern of spin waves. In a ferromagnet with DM interaction, the energy of the spin wave can be written in the following expression  \cite{Supplementary}: 

\begin{equation}
	\hbar\omega^c=D_{ex} q^2 - c D_{dm} q
    \label{eq:20}
\end{equation}
where $q$ is the wave vector, $c=\pm 1$, $D_{ex}$ is the spin wave stiffness, and $D_{dm}$ is the DM stiffness. The wave vector of a given magnon frequency can be as:

\begin{equation}
	q=\frac{c D_{dm}\pm\sqrt{D_{dm}^2\mp4D_{ex}\hbar \omega }}{2D_{ex}}
    \label{eq:21}
\end{equation}
In this ferromagnetic medium, spin waves propagate where their phase coherence leads to interference patterns. We consider a spherical wave $W_{ext}$, generated by two point sources whose intensity decays exponentially as:

\begin{equation}
	\mathbf{W}_{\mathrm{ext}}=W_0 \left( \frac{e^{i(q|\mathbf{r}-\mathbf{r}_1|-\omega t)}}{|\mathbf{r}-\mathbf{r}_1|} e^{|\mathbf{r}-\mathbf{r}_1|\xi}+\frac{e^{i(q|\mathbf{r}-\mathbf{r}_2|-\omega t)}}{|\mathbf{r}-\mathbf{r}_2|}e^{|\mathbf{r}-\mathbf{r}_2|\xi}\right)
    \label{eq:22}
\end{equation}
where $W_0$ is the amplitudes of the spherical waves, the coordinates of the vectors are given by:  $\mathbf{r}_1=(0,d+\eta)$, $\mathbf{r}_2=(0,\eta)$, $\mathbf{r}=(x,y)$, $\mathbf{r}-\mathbf{r}_1=(x, y-(d+\eta))$ and $\mathbf{r}-\mathbf{r}_2=(x, y-\eta)$ (the setup and the reference frame is shown in Supplementary Material, Section II, Fig.~S~3 \cite{Supplementary}), and $\xi$ is the attenuation coefficient induced by the medium where the spin waves propagate, which can be derived as:
\begin{equation}
	\xi^c=\frac{\alpha\omega}{2 \frac{D_{ex}}{\hbar}q-c \frac{D_{dm}}{\hbar}}
    \label{eq:23}
\end{equation}

For small angle deviations, we can approximate the averaged-square of angle deviation by:

\begin{equation}
\langle \theta \rangle^2\simeq \langle\frac{\gamma W_{ext}}{\omega} \rangle^2 + \sigma=\left(\frac{\gamma}{\omega}\right)^2\langle W_{ext}\rangle^2 + \sigma
\label{eq:24}
\end{equation}
where $\theta$ represents the average angle deviation of the moments from $z$-direction, $\gamma$ denotes gyromagnetic ratio, and $\sigma$ represents background spin intensity, potentially arising from microwave field-driven energy input.
All the derivations for Eqs.~\ref{eq:20}-\ref{eq:24} are provided in Supplementary Material, section I,II \cite{Supplementary}.

\begin{figure}[h]
    \centering    
\includegraphics[width=0.48\textwidth, height=0.3\textwidth]{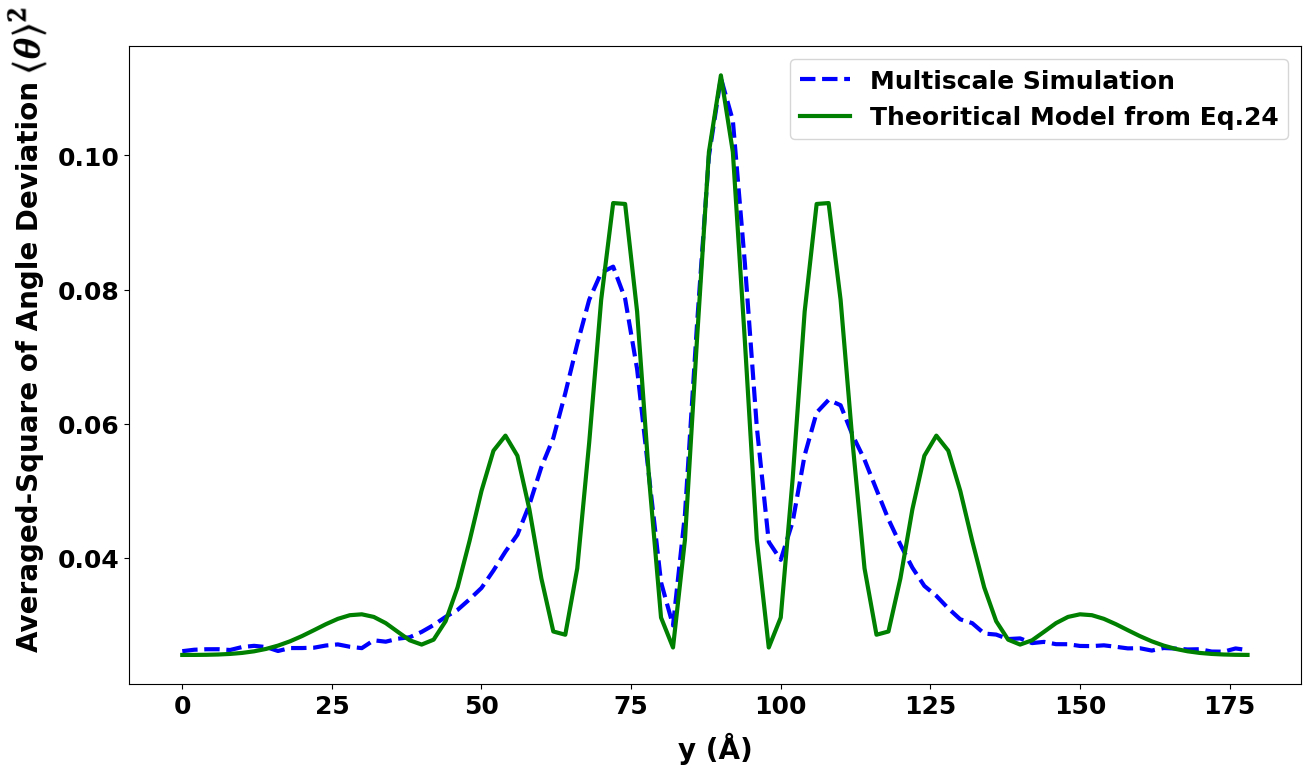}
    \caption{Comparison of micromagnetic simulation data and the theoretical derived formula which shows intereference pattern induced by averaged-square of angle deviation of the moment over time from $z$-direction, as a function of the length along the $y$-axis. The blue dashed line represent the micromagnetic data extracted from spin-wave propagation, while the green line corresponds to the theoretical derived formula given by Eq.~\ref{eq:24}.}
    \label{fig:3}
\end{figure}

This analogy suggests that spin-wave-mediated magnetization dynamics serve as a practical counterpart to the quantum mechanical double-slit experiment. As illustrated in Fig.~\ref{fig:3}, the fitted numerical results from Eq.~\ref{eq:24}, demonstrate that micromagnetic data follows the predicted theoretical form, indicating the wave nature of spin excitations in magnetic materials. The equation derived in this study effectively models the averaged-square of angle deviation of spin-wave interference in a micromagnetic system. The resemblance to the quantum double-slit experiment is evident through the presence of interference fringes modulated by an exponential decay factor.

\subsubsection{Domain Wall scattering and acceleration}

\begin{figure}[h]
    \centering    
    \includegraphics[width=0.45\textwidth, height=0.65\textwidth]{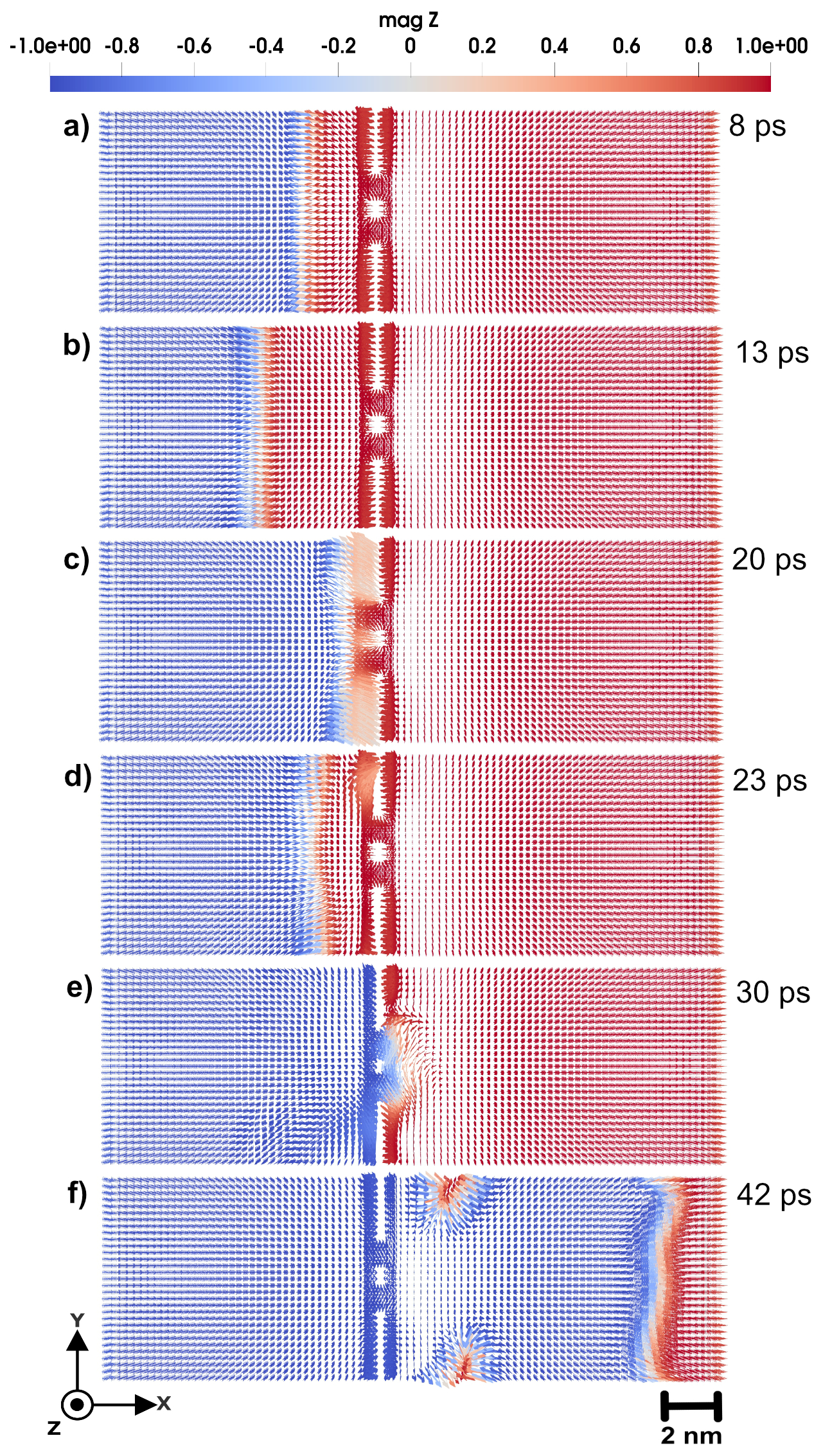}
    \caption{3D domain wall motion in the $x$-direction towards a double slit, induced by an applied external magnetic field. As the domain wall passes through the slits, it becomes perturbed, exhibiting a bouncing behavior for a period of approximately 25 ps. Panels a) to f) illustrate the evolution of the domain wall at different time frames, ranging from 8 to 42 ps. The color code shows the $z$ component of the normalized magnetization.}
    \label{fig:4}
\end{figure}

Double-slit configuration enables a focused analysis on how the domain wall structure and dynamics are influenced when passing through or interacting with geometrically confined defects of comparable scale. The 3D domain wall under investigation is a Néel-type wall with a width of approximately 3~\AA ($\Delta = \sqrt{\frac{A_{e}}{K_u}}$) (see e.g. Ref. \cite{Andersson2015}). The size of the atomistic region is $10 \times 60 \times 10$~\AA$^3$. Rectangular holes in this region have the size of $5 \times 20 \times 10$~\AA$^3$~  (upper and lower holes in Fig.~\ref{fig:4}) and $5 \times 6 \times 10$~\AA$^3$~ (the middle one in Fig.~\ref{fig:4}). The width of each slit is set to  $\sim$ 7~\AA~. This configuration enables a focused analysis of how the domain wall structure and dynamics are influenced when passing through or interacting with geometrically confined defects of comparable scale. The double-slit structure also serves as a controlled environment to study scattering effects and possible pinning mechanisms that arise due to spatially defined perturbations.

As shown in Fig.~\ref{fig:4}, where snapshots of magnetic configurations are taken in a time period up to 42 ps, the application of an external magnetic field of 3.0 T in the $z$-direction induces the movement of the domain wall from the left to the right side of the system. Upon approaching the double-slit structure located in the middle of the system (see Fig.~\ref{fig:4} a)-b)), the domain wall encounters reflection and is momentarily pushed back. Despite this initial repulsion, the domain wall continues its forward motion, only to be repelled a second time as it interacts with the slit edges. Eventually, the domain wall successfully passes through the double-slit structure. However, while passing the slits (Fig.~\ref{fig:4} c)-d)), the domain wall exhibits a pronounced curvature, arising from the spatial constraints imposed by the double-slit geometry and the localized magnetic interactions. This deformation underscores the influence of structural perturbations on domain wall dynamics, particularly in constrained environments. The domain wall motion under the applied magnetic field, offers a dynamic visualization of its behavior, including the moments of repulsion, re-approach, and eventual passing through the double slit. The dynamics of domain wall is presented in detail in Supplementary Material, Video 2 \cite{Supplementary}.

In the presence of a double-slit defect, the dynamics of domain wall motion exhibit significant changes compared to motion in a defect-free system. Specifically, when the domain wall passes through the double slit, it experiences a significant acceleration. Quantitatively, the post-slit velocity of the domain wall increases to approximately twice the velocity observed during propagation in the absence of the double slit. This phenomenon suggests that the geometric confinement and localized magnetic interactions induced by the double slit act to compress and reconfigure the domain wall structure, resulting in reduced pinning and enhanced mobility beyond the slit.

We developed a one-dimensional (1D) model provided in Supplementary Material, section III \cite{Supplementary}, which explains the reason for the increase in velocity and  explain the observed acceleration of magnetic domain walls as they traverse a double-slit geometry. To explain this behavior, we consider a 1D collective coordinate model describing the DW in terms of its center position $q(t)$, width $\Delta(t)$, and magnetization angle $\phi(t)$. The geometric constriction is modeled as a localized potential centered at the slit position $q^\prime$, described by:

\begin{equation}
	V(q-q\prime)=V_0 \; e^{-\frac{q-q\prime}{a}}
    \label{eq:26}
\end{equation}
where $a$ is the width of the potential and $V_{0}$ is the maximum value of potential.

This potential stores elastic energy due to domain wall compression. When deriving the equations of motion via the Lagrangian formalism, including dissipation effects, the resulting velocity expression for the domain wall is:

 \begin{align}
    \dot{q} &= \frac{1}{1+\alpha^2} \left( \alpha \gamma \Delta H - \alpha \Delta \eta e^{-\left(\frac{q-q'}{a}\right)^2} (q-q') - \alpha^2 \dot{\Delta} \ln\sqrt{2} \right. \nonumber \\
    &\quad \left. + \frac{\pi D \gamma}{4\Delta \mu_0 M_s} \cos\phi + \gamma \Delta H_k \sin(2\phi) \right)
    \label{eq:27}
\end{align}
where $\mu_0$ is the vacuum magnetic permeability, $M_s$ is the saturation magnetization, $D$ is the DM interaction, $K_0$ is the uniaxial anisotropy, $K$ is the second-order uniaxial transverse anisotropy, $H$ is an external magnetic field, $H_k=\frac{K}{\mu_0 M_s}$ is the uniaxial anisotropy field and $\eta=\frac{\gamma V_0}{\mu_0 M_s a^2}$.

Equation~\ref{eq:27} reveals the physical origins of acceleration. The chirality introduced by the DM interaction influences DW dynamics, with the direction and magnitude of its effect governed by the sign of the DM interaction constant $D$. A positive $D$ enhances DW velocity, while a negative $D$ tends to suppress it. Nonetheless, in the context of the double-slit geometry, the dominant mechanism responsible for the observed velocity increase stems from the interaction with the slit potential. Specifically, the second and third terms in Eq.~\ref{eq:27} play a central role.

The second term in Eq.~\ref{eq:27}, arising from the gradient of the localized potential, provides a forward driving force as the DW approaches the slit ($q - q\prime < 0$), resulting in an initial acceleration. Although this is followed by a decelerating effect once the wall passes the slit center ($q - q\prime > 0$), the structural modification imparted during the approach is critical. This interaction leads to a sudden compression of the DW, manifesting as a rapid decrease in the domain wall width ($\dot{\Delta} < 0$). As a result, the third term in Eq.~\ref{eq:27} becomes positive and contributes an additional boost to the DW velocity after it exits the slit.
This two-stage mechanism that consists of initial acceleration via geometric potential and post-slit acceleration through width contraction is responsible for the observed doubling of the velocity of the domain wall beyond the slit. Consequently, the double-slit structure effectively serves as a dynamic enhancer of domain wall speed, demonstrating a mechanism for controlling and tuning domain wall velocity through engineered nano-structures.

As a final comment to this section, it is worthy to mention that in order to detect interference patterns, the setup discussed in Fig.~\ref{fig:2} is much more likely to result in interesting results, as compared to that of Fig.~\ref{fig:4}. Hence, a driving field as describe by  Eq.~\ref{eq:19}, e.g. by electromagnetic field from a waveguide, together with the experimental setup according to Fig.~\ref{fig:2}, would allow, for example, magneto optical detection of a magnetic pattern that reflects magnonic interference phenomena. 

\subsection{Effects due to a tetrahedral anisotropy cluster}

Variation in magnetocrystalline anisotropy within a magnetic material can occur as a result of heterogeneity in its composition or can be artificially engineered through various processing methods. It is experimentally accessible to embed small crystalline regions (inclusions) within a magnetic host system, where these inclusions exhibit different magnetocrystalline anisotropy owing to distinct crystal orientations, phases, or compositions. Such heterogeneity leads to spatial variations in magnetic anisotropy magnitudes relative to the surrounding material. Experimentally, stress annealing is a widely used technique in Fe-based nanocrystalline alloys, where applying tensile stress during annealing induces magnetic anisotropy aligned with the stress direction by altering magnetoelastic interactions \cite{Ohnuma2005}. Moreover, hydrogenation of metallic multilayers such as Co/Pd interfaces can locally modulate magnetic anisotropy by altering the interfacial electronic structure, leading to spatially confined variations in anisotropy energy \cite{Klyukin2020}. To experimentally detect and map local magnetic anisotropy, techniques such as magnetic force microscopy (MFM), scanning Hall probe microscopy, and polarized neutron diffraction are employed \cite{Kibalin2019}. These allow for spatially resolved imaging of magnetic domains and anisotropy tensors, thereby revealing localized variations in magnetic behavior within heterogeneous or treated systems. Collectively, both material design and experimental treatments offer routes to achieve and characterize local magnetic anisotropy in metallic systems, enabling the development of advanced magnetic devices with region-specific magneto-crystalline anisotropies.

\begin{figure}[h]
    \centering
    \includegraphics[width=0.5\textwidth]{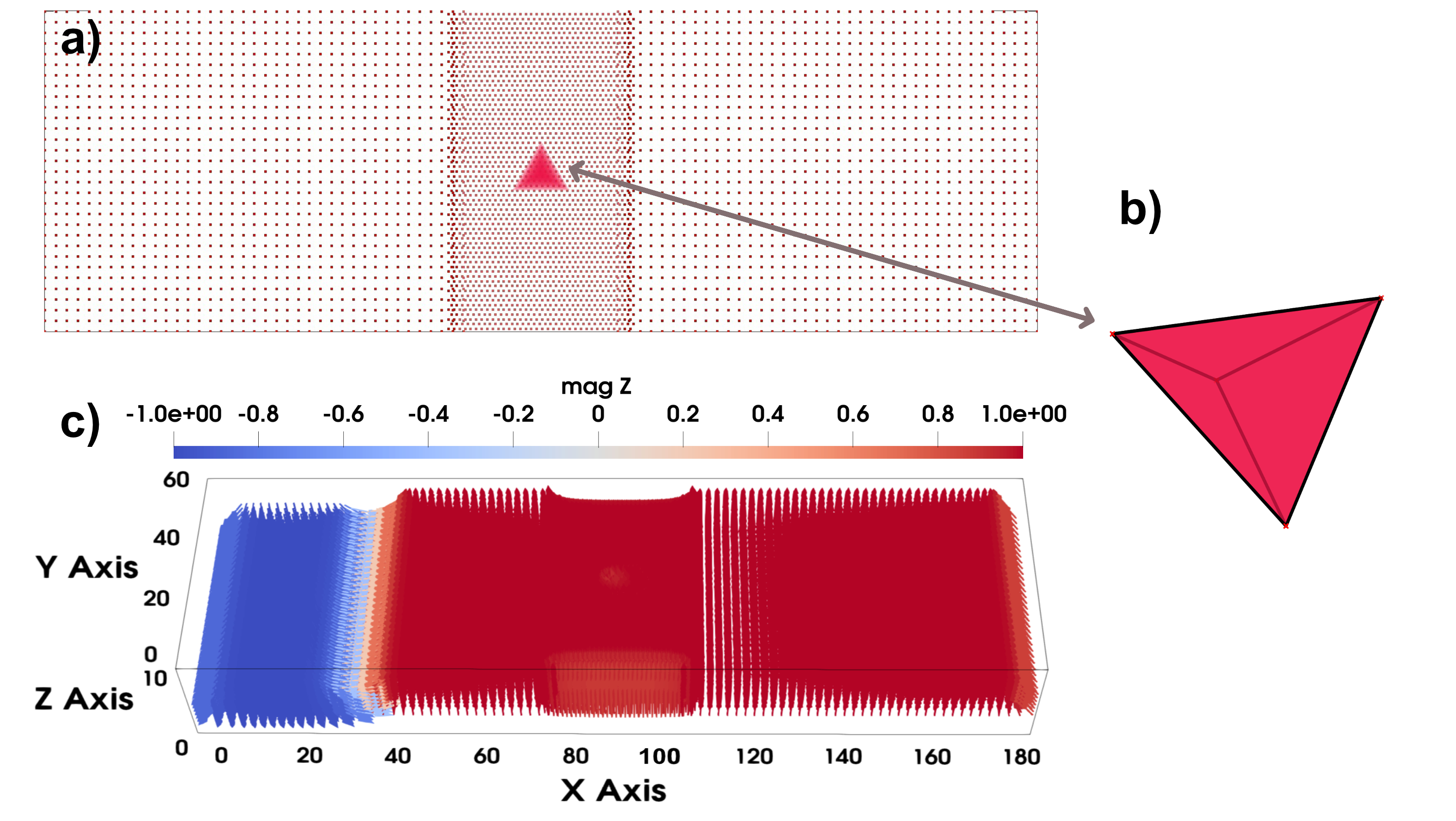}
    \caption{a) Top-view of the structure of the system consisting of a rectangular atomistic region in the middle that contains tetrahedron cluster at its center. 3b) A tilted view of the tetrahedron cluster. c) A 3D structure of a system including a 3D domain wall, with an atomistic region in the middle which shows a higher spin density in the atomistic region compared to the surrounding micromagnetic part. The color bar is showing the $z$ component of normalized magnetization.}
    \label{fig:5}
\end{figure}

Local lattice imperfections are also considered in this set of simulations. In particular, we consider a defect region with a tetrahedral geometry, as illustrated in Fig.~\ref{fig:5}. The micromagnetic system under investigation has dimensions of $180 \times 60 \times 10$~\AA$^3$. The atomistic region has the size of $30 \times 60 \times 10$~\AA$^3$. The tetrahedral cluster in the middle of the atomistic region contains approximately 720 atoms arranged in nine layers. Its overall dimensions are $10 \times 8.6 \times 8.1$~\AA$^3$. This cluster is assigned an anisotropy value different from the rest of the system, whereas all other simulation parameters are kept the same. Along the $y$-direction, the system has a periodic boundary condition. By systematically tuning the anisotropy of this local cluster, we illustrate its potential in influencing the dynamics of the domain wall and skyrmion.

\subsubsection{Domain Wall Pinning and Transformation}

The 3D domain wall under investigation here has a width of $\sim$ 3~\AA. By application of an external magnetic field in the negative $z$-direction, the domain wall starts to move from the left to the right side of the simulation box. In this study, we systematically tuned the uniaxial anisotropy of a smaller volume in the atomic region, to investigate its impact on the domain wall dynamics under two distinct conditions: i) with the local anisotropy aligned along the easy axis of the micromagnetic region (the $z$-axis of the simulation box) and ii) with local anisotropy aligned along the hard axes of the micromagnetic region (the $xy$-plane of the simulation box). The uniaxial anisotropy values explored for the local impurity were 0.11 mRy, 0.9 mRy, and 1.5 mRy. These values were chosen so that the first value (0.11 mRy) was very close to the anisotropy constant of the system (0.058 mRy) while the other ones were progressively increased in order to emphasize the influence of the tetrahedron anisotropy as compared with the host material. The outcomes of the simulations reveal a complex interplay between the anisotropy strength, its directional alignment, and the domain wall behavior as it interacts with the tetrahedral region. This approach offers a pathway to investigate how local anisotropy variations can alter domain wall properties such as width, pinning potential, and mobility.

\begin{figure}[h]
    \centering
    \includegraphics[width=0.45\textwidth, height=0.65\textwidth]{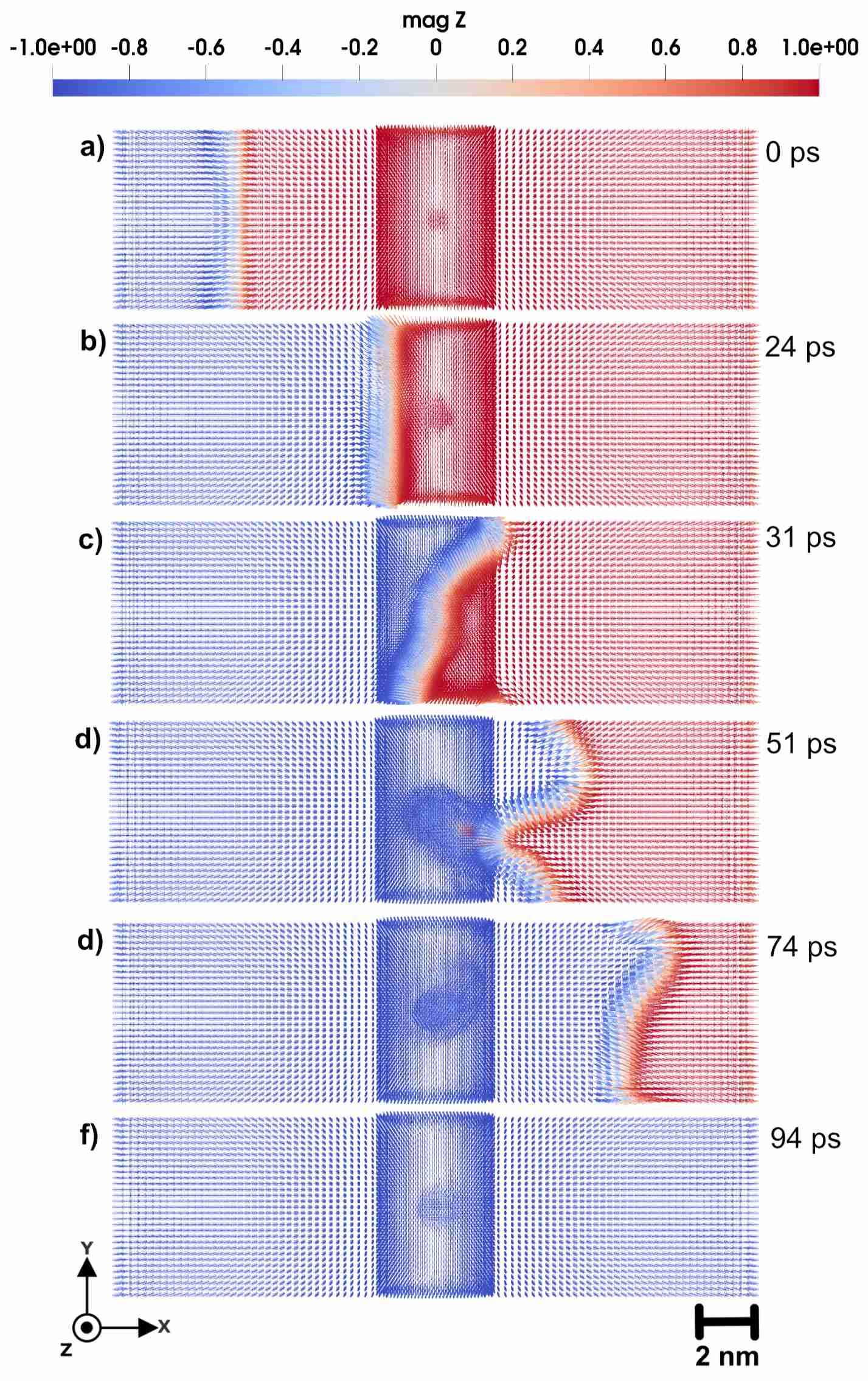}

    \caption{3D Domain wall motion by applying magnetic field of 0.47 T in negative $z$-direction. In the center of the atomistic region (shown by slightly deeper color) a local cluster is placed that has uniaxial anisotropy with hard axis along the $z$-direction, with strength of 0.11 mRy.  The color bar is showing the $z$ component of normalized magnetization.}
    \label{fig:6}
\end{figure}

We first considered a uniaxial anisotropy of the defect with the tetrahedral region set to 0.11 mRy. We investigate two directions for the anisotropy:  along the easy ($z$-direction) and hard axis ($x$-$y$ plane).  In Fig.~\ref{fig:6}, the tetrahedron has anisotropy along hard axis ($x$-$y$ plane). By applying a magnetic field of 0.47 T in negative $z$-direction, the domain wall successfully passes from left to right in the simulation cell, as shown in Fig.~\ref{fig:6} a)-f). However, due to the influence of the tetrahedron topology and magnetic properties, the domain wall experiences slight disturbances, manifesting as localized deformations and pinning during the time evolution. This phenomenon occurs when the local anisotropy is aligned along the easy axis of the micromagnetic region, as well as when the local anisotropy is along the hard axis of the micromagnetic region, as illustrated in Supplementary Material, Video 3,4 \cite{Supplementary}.

\begin{figure}[h]
    \centering
     \includegraphics[width=0.45\textwidth, height=0.65\textwidth]{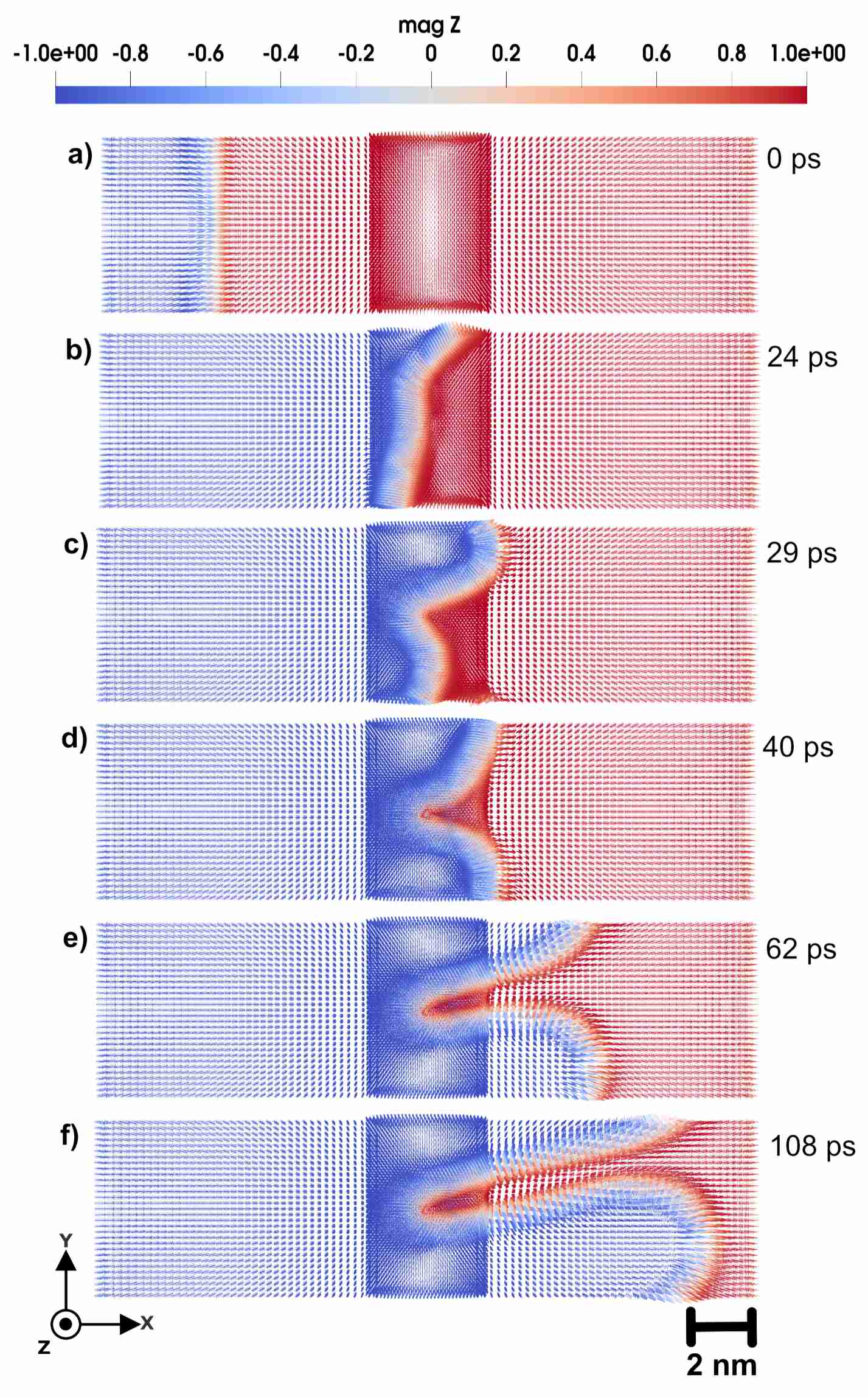}
    \caption{3D Domain wall motion driven by applying a magnetic field of 0.47 T in negative $z$-direction. The domain wall is seen to move through a tetrahedron cluster with uniaxial anisotropy that has easy axis along $z$-direction and strength of 0.9 mRy. Simulations that consider a local anisotropy that has the $z$-axis as a hard axis produces similar results. The color bar shows the $z$ component of the normalized magnetization.}
    \label{fig:7}
\end{figure}

These findings can be directly linked to the microscopic origin of the Barkhausen effect, as previously discussed in Ref.~\cite{Mendez2020}. In the present study, the tetrahedral impurity region serves as a localized anisotropy-induced defect, which behaves similarly to irregularities in the lattice such as dislocations or impurities. As the domain wall interacts with this region, it experiences localized pinning, where components of the wall become temporarily stuck due to the strong energy barrier associated with anisotropy. Similar to the behavior illustrated in earlier studies where parts of the domain wall are trapped by defects, the pinned region moves slowly behind the rest of the wall. The following unpinning phenomena, especially under increased magnetic field strength, are similar to the sudden jumps in magnetization observed experimentally in the Barkhausen effect, highlighting how atomic-scale anisotropy variations can lead to abrupt, collective magnetic changes. The atomistic modeling approach adopted here provides a deeper insight into such phenomena, capturing the complex interplay of local magnetic parameters at the atomic level.

As the uniaxial anisotropy of the impurity region is increased to 0.9 mRy, the interaction with the domain wall  becomes more pronounced, as one can see in Fig.~\ref{fig:7}. By applying an external field of 0.47 T in negative $z$-direction, the domain wall moves from left to right in the simulation cell. Upon approaching the impurity region, the domain wall becomes pinned and takes a significantly longer time to pass the region. In fact, as Fig.~\ref{fig:7} f) shows, during the time of the simulations the domain wall has a part that never leaves the impurity region. This behavior occurs for the both cases when anisotropy is along easy and hard axis of the micromagnetic region as shown in Supplementary Material, Video 5, 6 \cite{Supplementary}. This is attributed to the stronger anisotropy-induced energy barrier, which opposes to the domain wall motion. 

\begin{figure}[h]
    \centering
   \includegraphics[width=0.44\textwidth, height=0.8\textwidth]{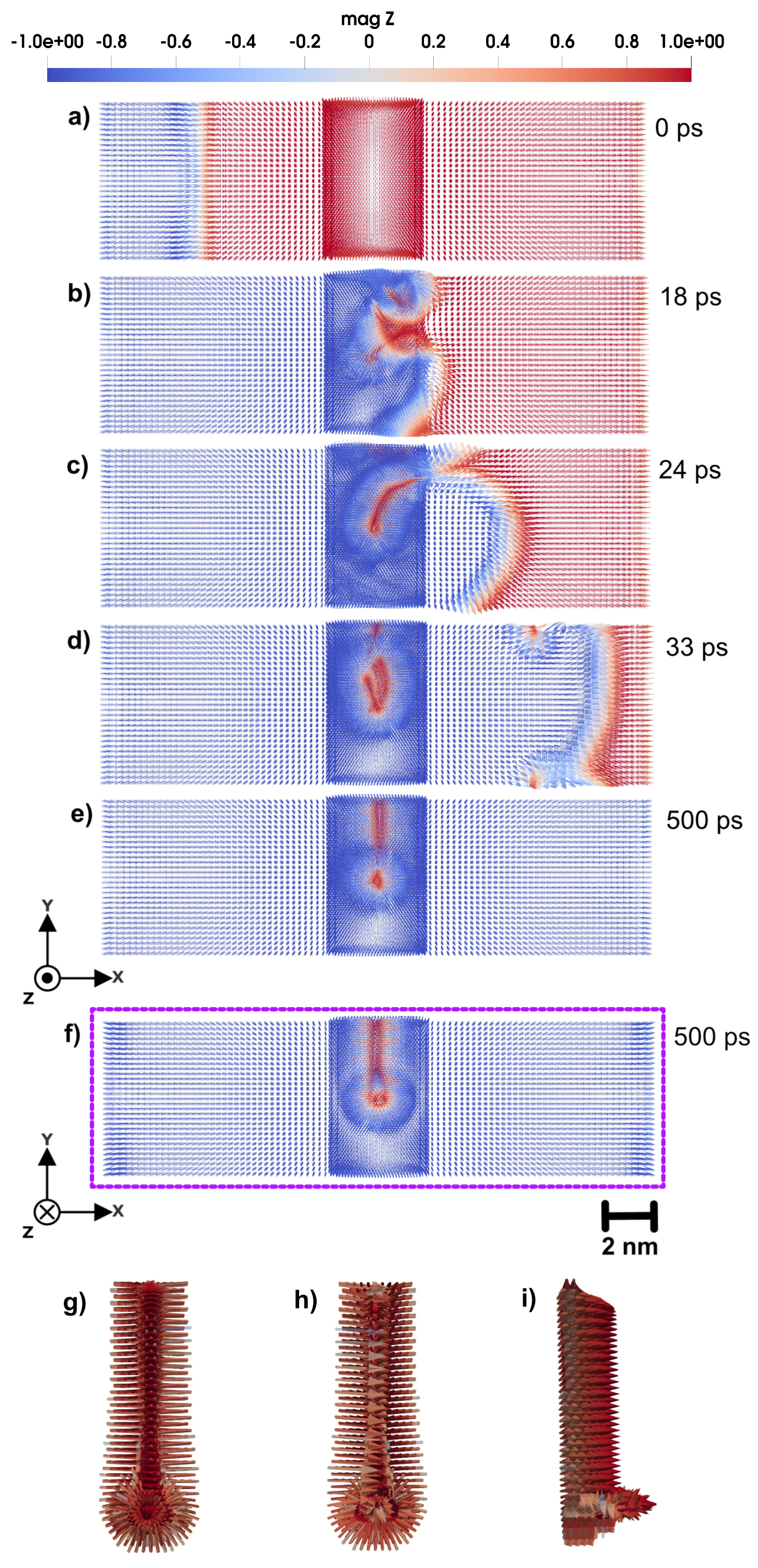}
    \caption{3D Domain wall motion by applying a magnetic field of 1.5~T so that a domain wall moves through the tetrahedron cluster with uniaxial anisotropy that has easy axis along $z$-direction and strength of 0.9 mRy. Panels a) to e) represent a top view of the magnetic texture along the $z$-direction, while panel f) provides a bottom view along the same direction, elucidating the shape of the created skyrmion in three dimensions. Panels g), h), and i) respectively illustrate the bottom, top, and lateral views of the 3D skyrmion, each exhibiting a 90-degree bend. The color bar is showing the $z$ component of normalized magnetization.}
    \label{fig:8}
\end{figure}

\begin{figure}[h]
    \centering
   \includegraphics[width=0.44\textwidth, height=0.8\textwidth]{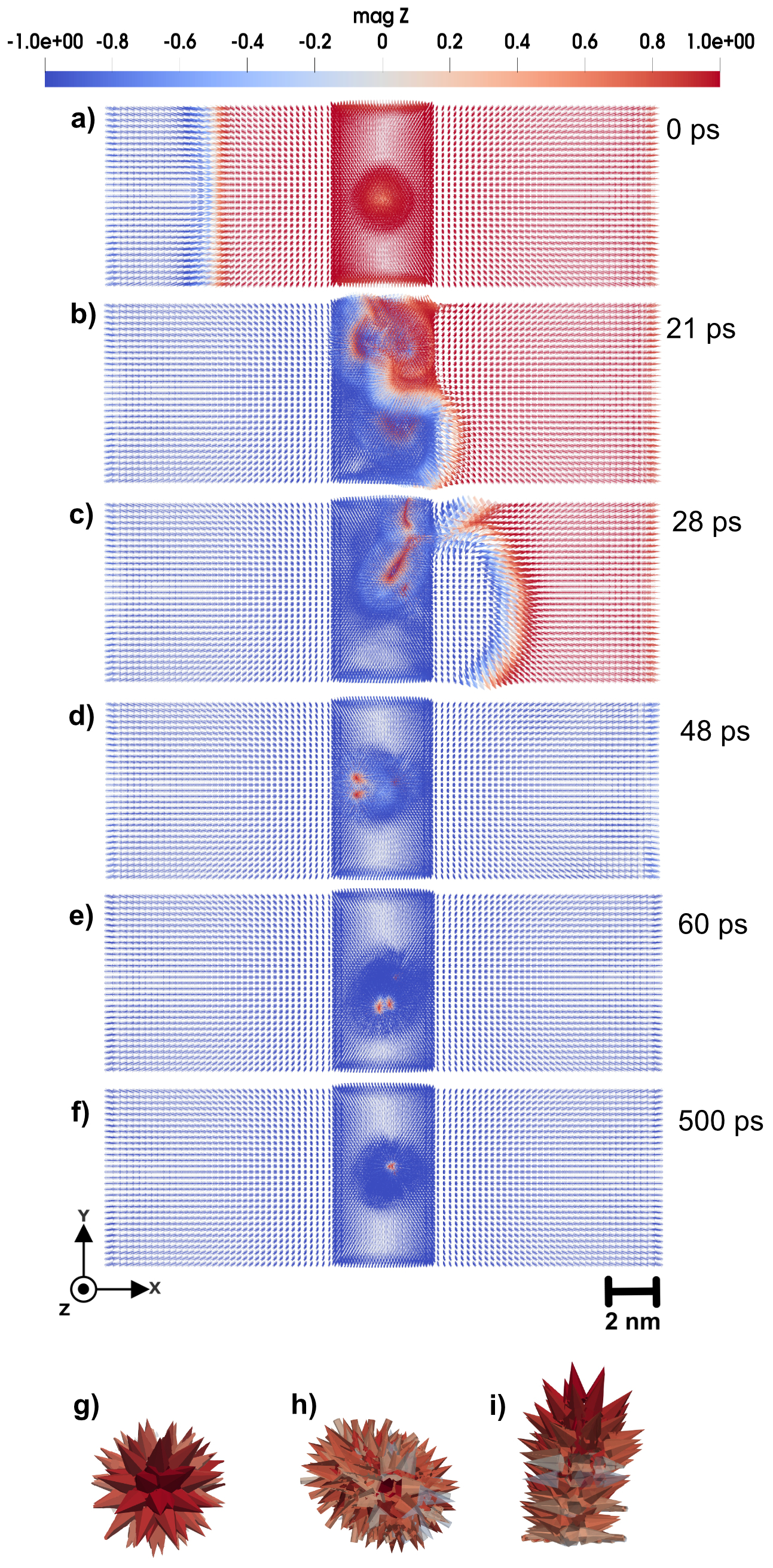}
    \caption{3D domain wall motion by applying a magnetic field of 1.5~T through a system with a tetrahedron cluster with uniaxial anisotropy, where the $z$-direction serves as the hard axis with a strength of 0.9 mRy. Panels a) to f) display the time evolution of the domain wall at different times. Panels g), h), and i) illustrate a generated hedgehog skyrmion from the top, bottom, and side views, respectively. The color bar is showing the $z$ component of normalized magnetization.}
    \label{fig:9}
\end{figure}

When the applied field is increased to 1.5~ T, the domain wall can overcome the impurity region after being pinned to it but the domain wall exhibits dramatically different behaviors depending on whether the anisotropy is aligned along the easy or hard axis of the micromagnetic region. In the case where the impurity uniaxial anisotropy is along the easy axis of the micromagnetic region, an external magnetic field of 1.5~T in negative $z$-direction causes the domain wall to approach the impurity region, where begins to lose its coherent structure (see Fig.~\ref{fig:8}). This leads to the emergence of stable tubular configurations, as shown in Fig.~\ref{fig:8} e),f). This tubular structure persists and stabilizes, suggesting that the strong easy-axis anisotropy introduces a localized energy landscape that fundamentally alters the domain wall configuration. The dynamics of this domain wall is presented in Supplementary Material, Video 7 \cite{Supplementary}. Interestingly, the domain wall dynamcis of Fig.~\ref{fig:8} serves as an initial step for skyrmion formation, and in this scenario, the generated 3D skyrmion exhibits a 90-degree bending. This is illustrated in  Fig.~\ref{fig:8} g)- i), where different projections of the skyrmion structure are illustrated.

\begin{figure}[h]
    \centering
   \includegraphics[width=0.44\textwidth, height=0.8\textwidth]{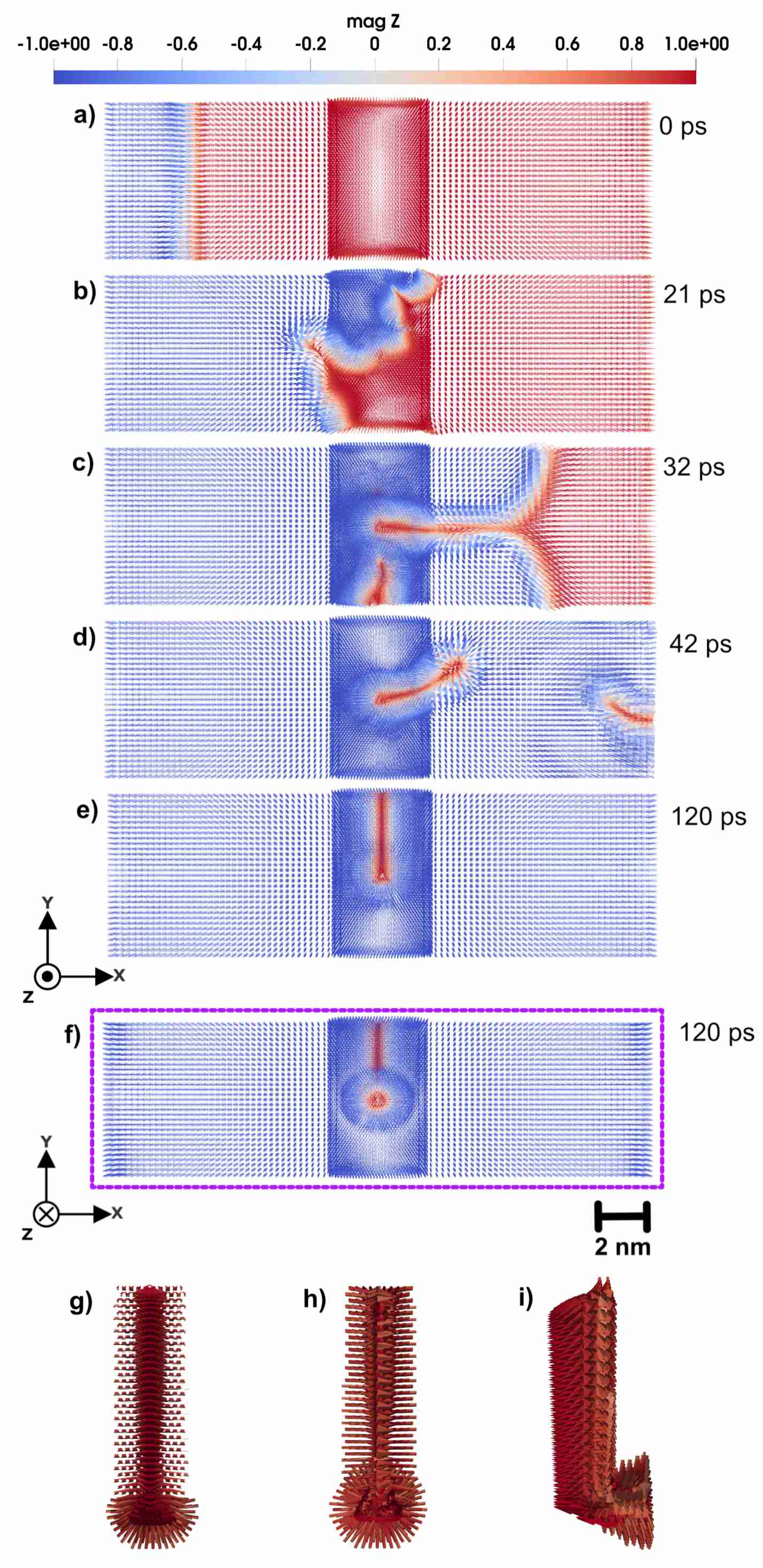}
    \caption{3D Domain wall motion by applying a magnetic field of 2~T so that a domain wall moves through the tetrahedron cluster with uniaxial anisotropy that has easy axis along $z$-direction and strength of 1.5 mRy. Panels a) to e) represent a top view of the magnetic texture along the $z$-direction, while panel f) provides a bottom view along the same direction, elucidating the shape of the created skyrmion in three dimensions. Panels g), h), and i) respectively illustrate the bottom, top, and lateral views of the 3D skyrmion, each exhibiting a 90-degree bend. The color bar is showing the $z$ component of normalized magnetization.}
    \label{fig:10}
\end{figure}

\begin{figure}[h]
    \centering
  \includegraphics[width=0.43\textwidth, height=0.75\textwidth]{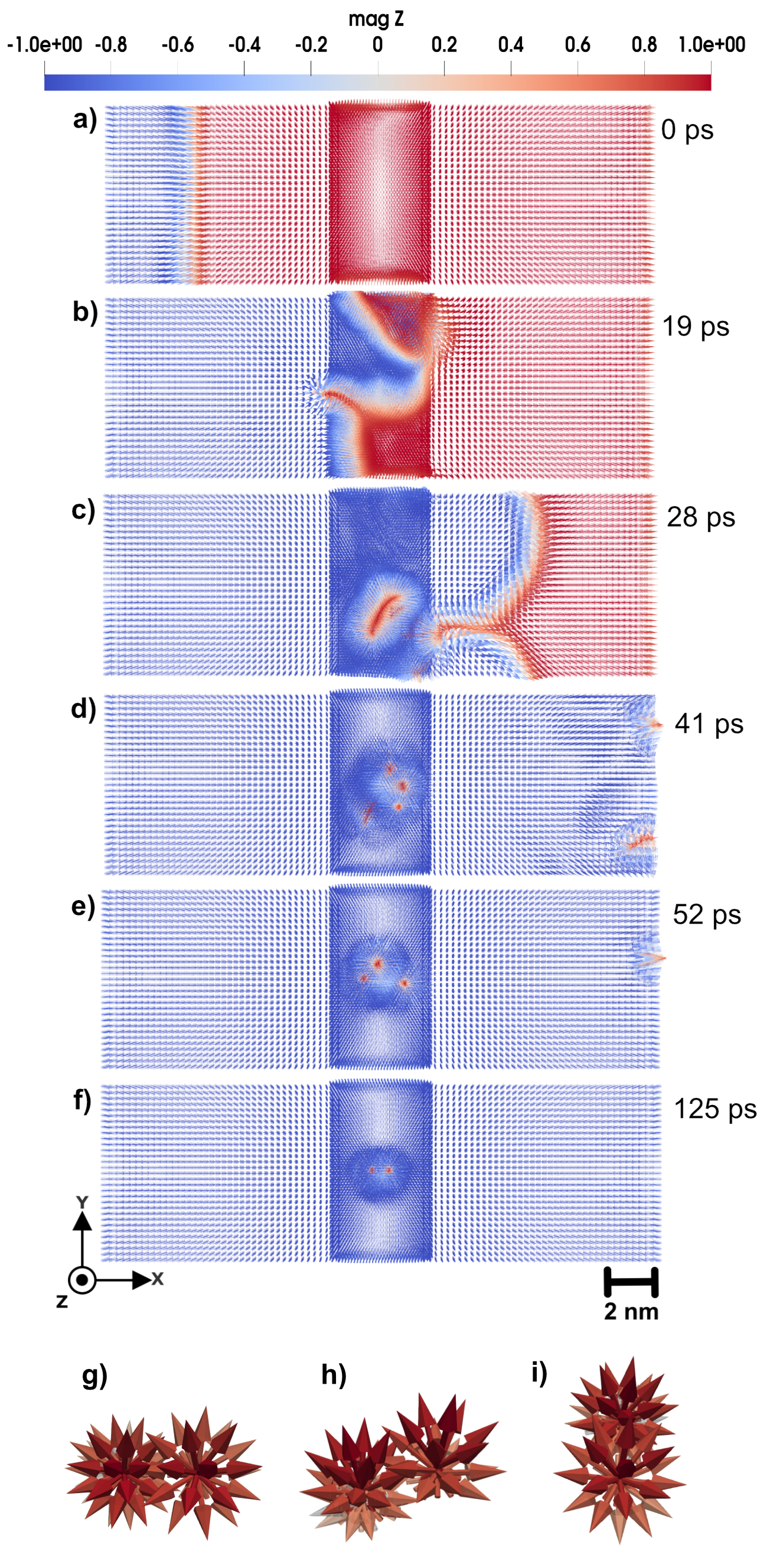}
    \caption{3D domain wall motion by applying a magnetic field of 2~T through a system with a tetrahedron cluster with uniaxial anisotropy, where the $z$-direction serves as the hard axis with a strength of 1.5 mRy. Panels a) to f) display the time evolution of the domain wall at different times. Panels g), h), and i) illustrate three generated hedgehog skyrmions from the top, bottom, and side views, respectively. The color bar is showing the $z$ component of normalized magnetization.}
    \label{fig:11}
\end{figure}

The 90-degree bent skyrmion has potential applications in magnon transport.
Given this unique curvature, it is plausible that magnons could be excited at the bottom of the system and subsequently guided along the bent domain wall, effectively channeling spin waves through a non-linear path. Such a configuration could serve as a waveguide for magnons, enabling controlled spin wave propagation and bending, that could become a crucial aspect for future magnonic circuits and information carrying technologies \cite{Chumak2015, Kruglyak2010, Garst2017, Khitun2010}.

By having uniaxial anisotropy of the defect region aligned along the hard axis of the micromagnetic region with a strength of 0.9 mRy, and applying field of 1.5~T, the domain wall also loses its coherence upon entering the tetrahedral region, as presented in Fig.~\ref{fig:9} and Supplementary Material, Video 8 \cite{Supplementary}.The disturbance results in the formation of a tubular hedgehog skyrmion configuration shown in Fig.~\ref{fig:9} g)-i). This structure remains stable after its formation, highlighting a distinct anisotropy-dependent response.

The triangular tetrahedral cluster appears to play a role in this phenomenon by influencing the spin wave dynamics. If spin waves are pumped at the bottom, they may propagate along the structure, demonstrating controlled directional magnon transport. This cluster may function as a magnon conduit, absorbing spin waves in the central region and redirecting them towards the edges of the system. Such an effect could enable localized magnon control, which is essential for magnonic computing and low-power spintronic applications \cite{Serga2010, Krawczyk2014,Chumak2015}. If validated through experimental studies, this effect could lead to reconfigurable magnonic waveguides, where skyrmions guide spin waves \cite{Zhang2015}, efficient magnon transmission in curved geometries, minimizing energy loss \cite{Streubel2016}, and spin-wave-based logic elements, where anisotropy tuning controls directional magnon flow \cite{Nagaosa2013}. This behavior suggests that localized anisotropy variations can induce complex spin textures, potentially leading to the formation of deformed or bent skyrmions in 3D magnetic systems \cite{Mendez2020}.

 With the anisotropy further increased to a high value of 1.5 mRy, the domain wall dynamics remain qualitatively similar to those observed at 0.9 mRy, by applying a 2~T external magnetic field in the negative $z$-direction. When the uniaxial anisotropy within the tetrahedral region is aligned with the easy axis of the surrounding micromagnetic region, applying field drives the domain wall toward the impurity and begins to lose its coherent structure (see Fig.~\ref{fig:10}). The tubular structure illustrated in Fig.~\ref{fig:10}g)-i), remains stable over time. This configuration indicates that in the same case as previous shown data, the strong easy-axis anisotropy creates a localized energy landscape that significantly modifies the domain wall structure. The detailed evolution of this domain wall can be seen in Supplementary Material, Video 9 \cite{Supplementary}.

When the uniaxial anisotropy within the defect region is oriented along the hard axis of the surrounding micromagnetic region, the domain wall again loses its coherence upon entering the tetrahedral region, as shown in Fig.~\ref{fig:11} and Supplementary Material, Video 10 \cite{Supplementary}. This deformation gives rise to three distinct hedgehog structures(see Fig.~\ref{fig:11} g)-i)). 

When the domain wall passes through the defect, we observe clear changes in the skyrmion number depending on both the strength and direction of the magneto-crystalline anisotropy of tetrahedron. Considering the anisotropy of the tetrahedron to be equal to 0.9~mRy, when anisotropy is along the easy axis of the micromagnetic region, the skyrmion number reaches a value of $Q^V=$~–40. But when the anisotropy is along the hard axis, the value drops down to $Q^V=$~–26. Increasing the anisotropy of the tetrahedron to 1.5~mRy in the easy axis direction, it leads a bit to an increase in the skyrmion number, reaching $Q^V=$~–46 with respect to the case with $K_a=0.9$~ mRy. Interestingly, when the anisotropy constant is 1.5~mRy along the hard axis, the value stays $Q^V=$~–26, showing no change. The higher order skyrmions generated in the current setup provides a clear indication about the complex magnetic texture shown in Figs.~\ref{fig:8}-\ref{fig:11} g)-i). 

These results suggest that easy-axis anisotropy tends to favor more twisting in the spin texture as the domain wall interacts with the defect, likely because it supports out-of-plane alignment of spins. That allows for more complex, topologically rich structures to form, around the distorted region introduced by the tetrahedron. On the other hand, hard-axis anisotropy seems to suppress that kind of twisting, so the topological charge stays lower and does not vary much even when the anisotropy strength is increased. It appears that once anisotropy reaches a level where spin canting is strongly suppressed, further enhancement has negligible influence. These observations underscore the significant role of localized anisotropy variations in dictating the behavior of magnetic textures such as domain walls and formation of topological structure.

\subsubsection{Skyrmion-Defect Interactions}

 \begin{figure}[h]
    \centering
  \includegraphics[width=0.45\textwidth, height=0.6\textwidth]{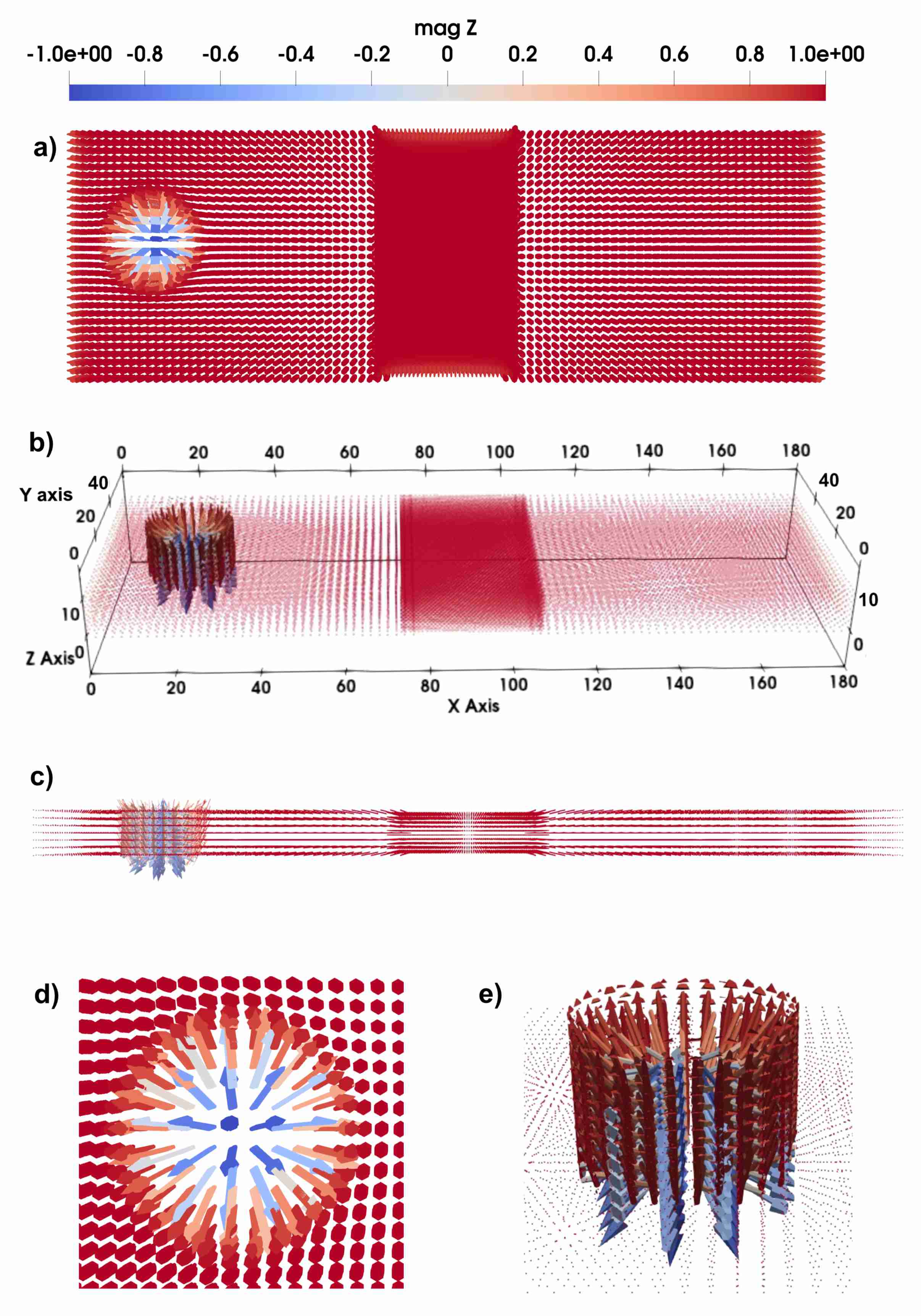}
    \caption{The magnetic texture of a 3D skyrmion. From panels a) to e), various perspectives of the 3D skyrmion are presented, highlighting its tubular magnetic configuration, which resembles a Neél-type structure observed in 2D magnetic systems. Panel a) represents a top view, whereas panels b) and c) provide a lateral view. Panels d) and e) show the skyrmion magnified, providing detailed information about the magnetic texture. The color bar is showing the $z$ component of normalized magnetization.}
    \label{fig:12}
\end{figure}

A 3D skyrmion is generated in the micromagnetic region, by applying a local magnetic field in the negative $z$-direction as an external excitation. We used a 300 T local field to tilt the direction of the local magnetic moments in the opposite direction of the neighboring atoms with the aim to generate the 3D skyrmion quickly and prevent a time consuming simulation. This field modulates the local magnetization, leading to the formation of a tubular skyrmion in 3D, a topologically stable structure characterized by a swirling magnetization pattern along the depth of the system as is illustrated in Fig.~\ref{fig:12}.

Once the skyrmion is generated and stabilized, it can be moved by applying a spin transfer torque (STT), a phenomenon where the angular momentum of a spin-polarized current influences the local magnetization dynamics. The STT effect is included in the LLG equation of the micromagnetic region, via an additional adiabatic term \cite{Schieback2007, Zhang2004}:

\begin{equation}
\mathbf{H}_{STT} = -(\mathbf{u} \cdot \nabla)\mathbf{m}
\label{eq:28}
\end{equation}
where $\mathbf{H}_{STT}$ is the STT effective field,  the vector $u$ is expressed in velocity units and is directly proportional to the applied current velocity:

\begin{equation}
\mathbf{u} = \frac{Pg\mu_{B}}{2eM_s}\mathbf{j}_{e}
\label{eq:29}
\end{equation}
where $P$ is polarization, $g$ is the Lande factor, $j_e$ is electron current density, $e$ is the electron charge and $M_s$ is saturation magnetization.

\begin{figure}[h]
    \centering
\includegraphics[width=0.44\textwidth, height=0.7\textwidth]{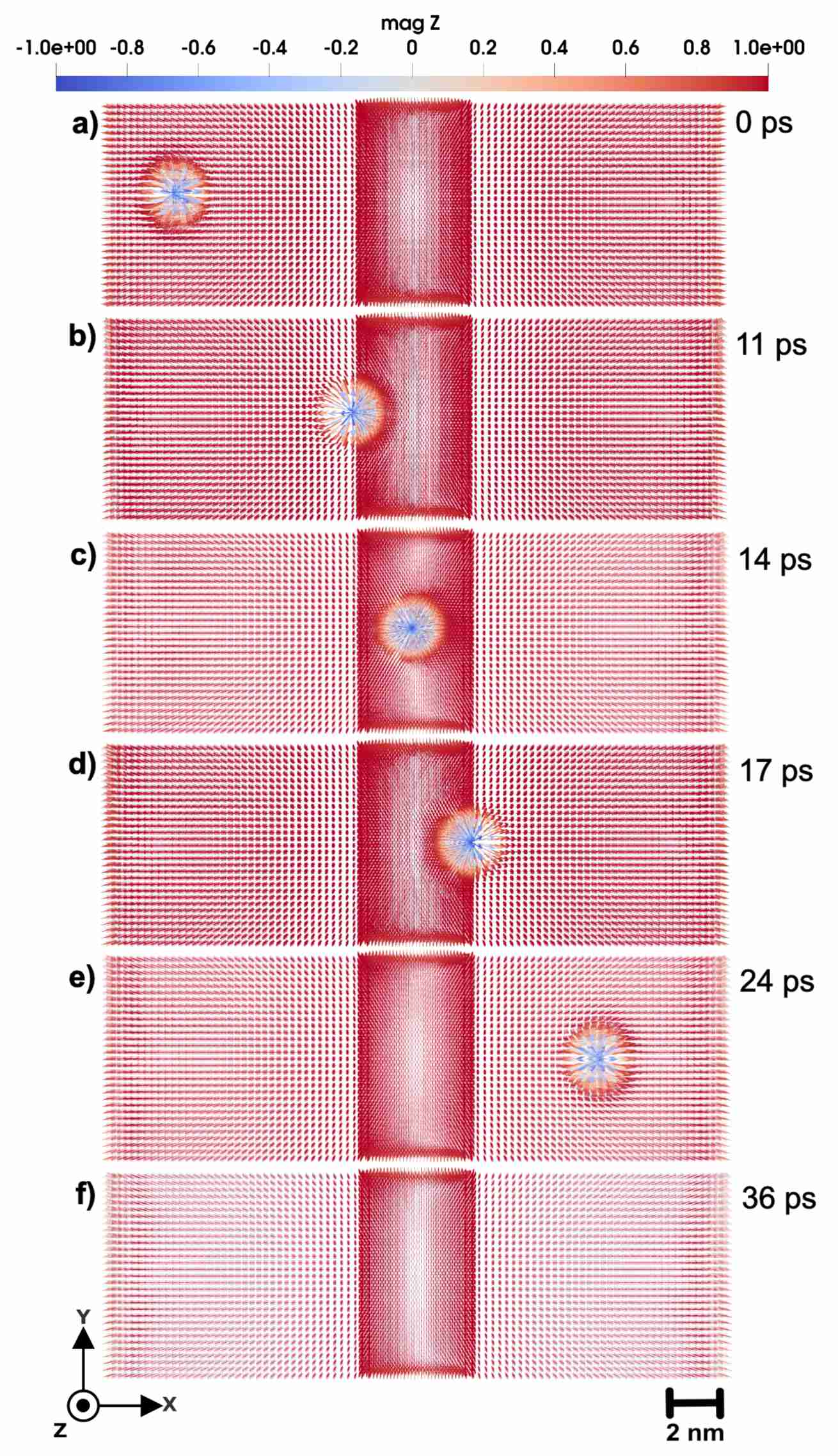}
    \caption{3D Skyrmion motion by applying a STT of 30~m/s through the simulation cell, with a defect region provided by a tetrahedron cluster with uniaxial anisotropy. The easy axis is along the $z$-direction and the uniaxial anisotropy constant is 0.11 mRy. The color bar is showing the $z$ component of normalized magnetization.}
    \label{fig:13}
\end{figure}

\begin{figure}[h]
    \centering
\includegraphics[width=0.44\textwidth, height=0.7\textwidth]{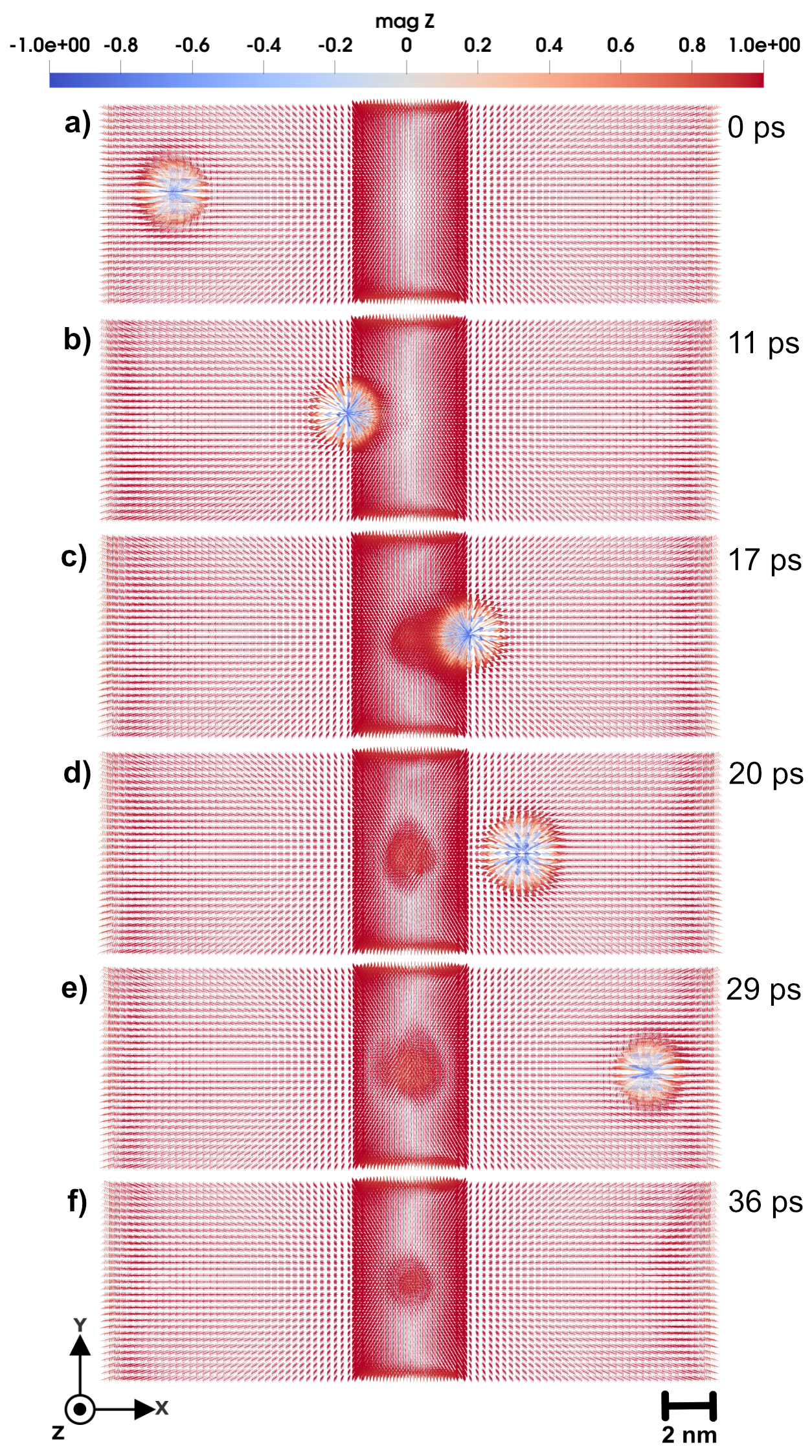}
    \caption{3D Skyrmion motion by applying a STT of 30~m/s through the simulation box, with a defect region provided by a tetrahedron cluster with uniaxial anisotropy. The hard axis is along the $z$-direction and the uniaxial anisotropy constant is 0.11 mRy. Similarly, a simulation with hard axis anisotropy produces the same result. The color bar is showing the $z$ component of normalized magnetization.}
    \label{fig:14}
\end{figure}

In this section, we explore the behavior of a 3D skyrmion as it interacts with a tetrahedral region featuring a change in the anisotropy values. The skyrmion motion is driven by STT, that is induced by a spin current applied from the left of the simulation cell, causing the skyrmion to move towards the right. The velocity of the skyrmion due to the STT depends on the parameters of Eq.~\ref{eq:29}, and is tuned to ensure the controlled and observable motion. As the skyrmion approaches to the tetrahedral region, its interaction dynamics are analyzed under different anisotropy strengths and orientations (easy or hard axis). The skyrmion response is significantly different, depending on the easy axis orientation of the local defect region. With the tetrahedral anisotropy set to 0.11 mRy and aligned along the easy axis, the skyrmion, driven by a STT with velocity $30~m/s$, traverses the region without notable deformation or disruption as shown in Fig.~\ref{fig:13}, and Supplementary Material, Video 11 \cite{Supplementary}.  When the anisotropy is instead aligned along the hard axis, the skyrmion still passes through the tetrahedral region(see Fig.~\ref{fig:14}), however, exhibiting a temporary and small size increase (diameter increases $\approx$ 6~\AA) during and shortly after the interaction. The dynamics of this skyrmion is presented in Supplementary Material, Video 12 \cite{Supplementary}.

\begin{figure}[h]
    \centering
\includegraphics[width=0.44\textwidth, height=0.7\textwidth]{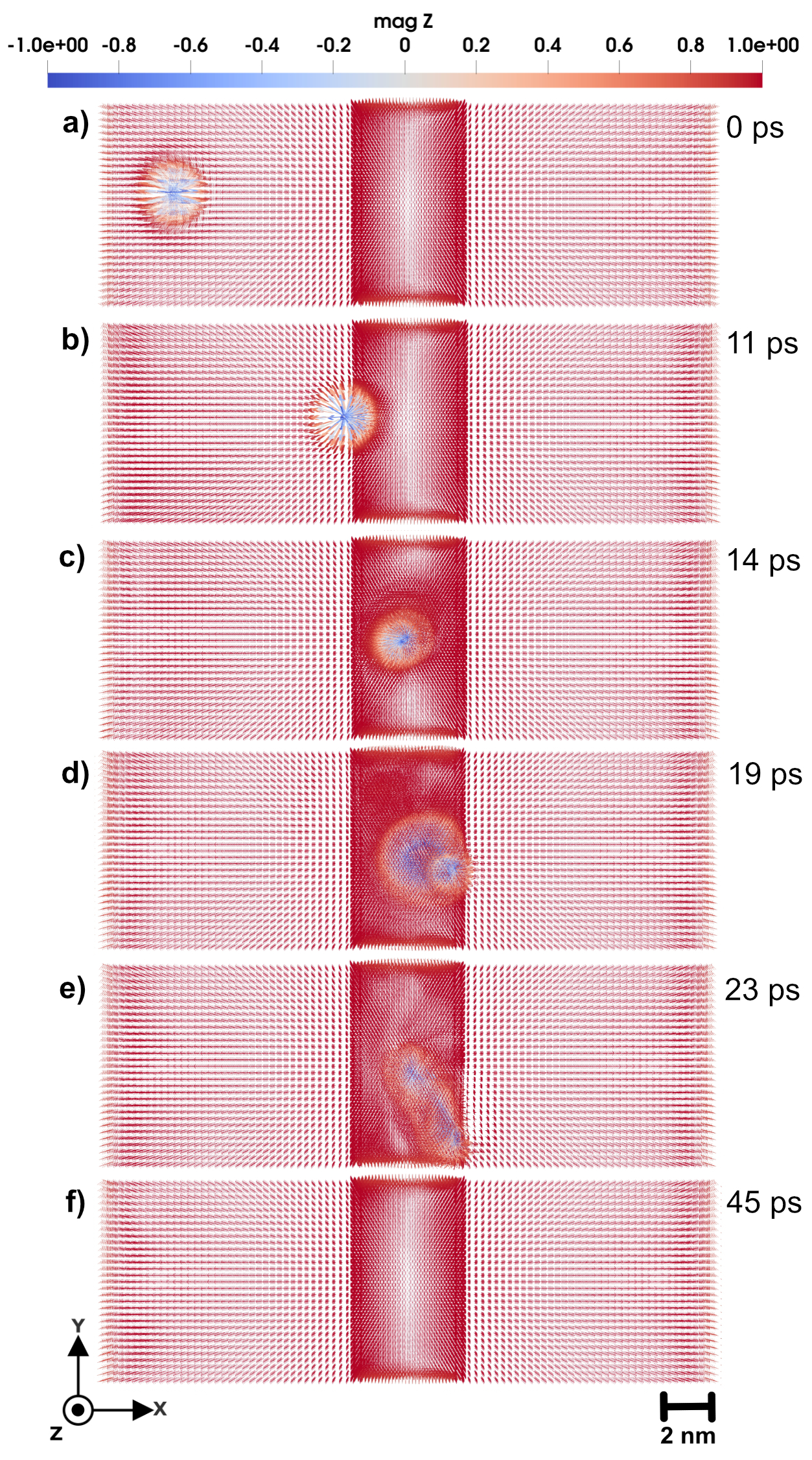}
    \caption{3D Skyrmion motion by applying STT of 30~m/s through the simulation cell, with a defect region shaped as a tetrahedron cluster with uniaxial anisotropy and easy axis along $z$-direction with strength of 0.9 mRy. The color bar is showing the $z$ component of normalized magnetization.}
    \label{fig:15}
\end{figure}

Unlike domain walls, which exhibit curvature and pinning, due to the interaction provided by the tetrahedral region, the skyrmion maintains its structural integrity. This resilience is attributed to the inherent stability provided by the skyrmion topology, which enables it to navigate low-anisotropy regions with minimal perturbation. At a higher anisotropy of 0.9 mRy and with a STT applied along the $x$-direction with a velocity of $30~m/s$, the skyrmion motion through the tetrahedral region slows down. When the anisotropy of the cluster is aligned along the easy axis of micromagnetic region, the moving skyrmion disappears when it reaches the tetrahedral region, as illustrated in Fig.~\ref{fig:15} and Supplementary Material, Video 13 \cite{Supplementary}. This disappearance suggests that the strong easy-axis anisotropy creates a local energy barrier landscape that destabilizes the skyrmion core, effectively destroying its topological structure.

With the anisotropy of the tetrahedron aligned along the hard axis, the skyrmion again successfully passes the tetrahedral region, however, it has an increase in its size as shown in Fig.~\ref{fig:16}. The dynamics of this skyrmion is presented in Supplementary Material, Video 14 \cite{Supplementary}.

\begin{figure}[h]
    \centering
\includegraphics[width=0.44\textwidth, height=0.7\textwidth]{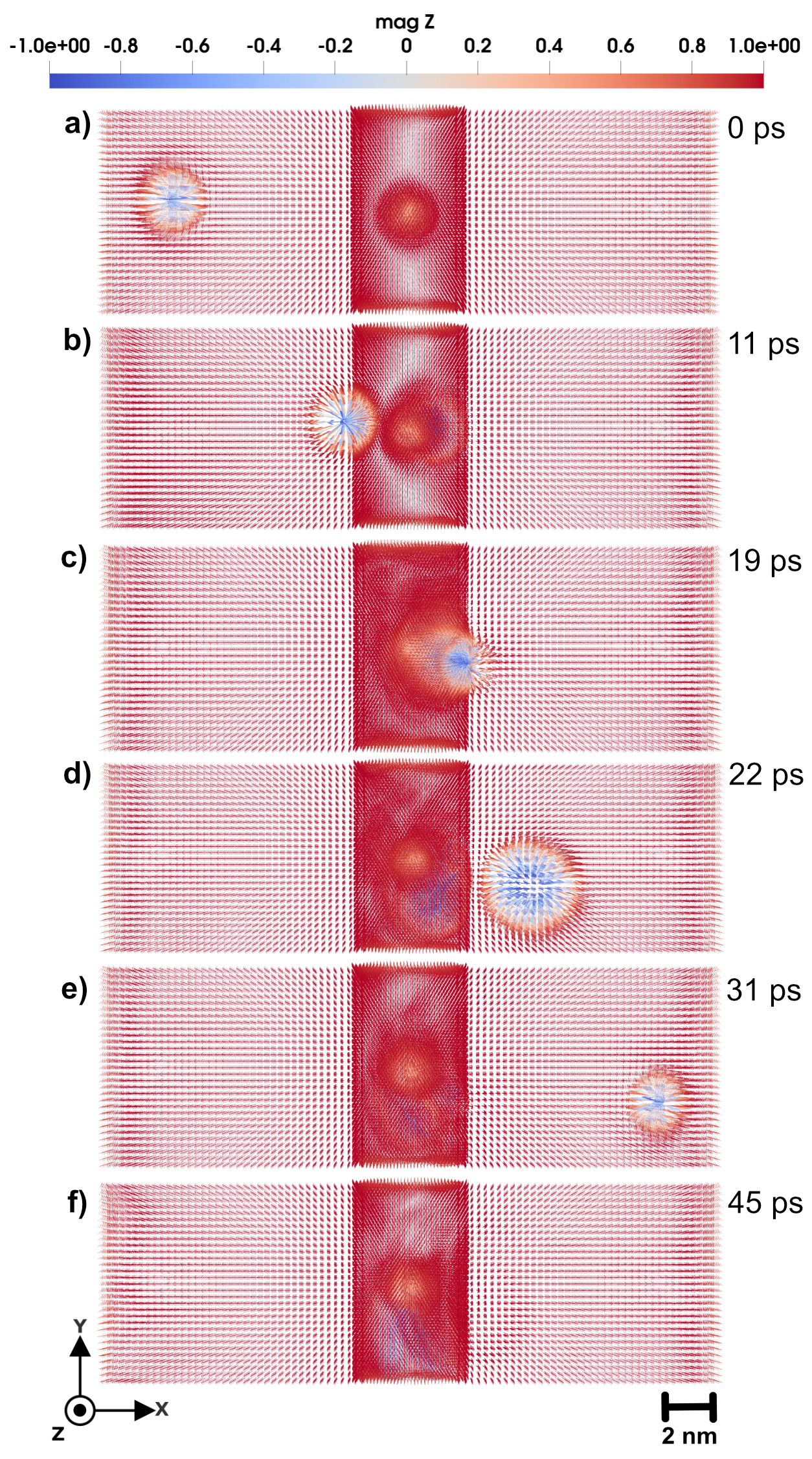}
    \caption{3D Skyrmion motion by applying STT of 30~m/s through the simulation cell, with a defect region shaped as a tetrahedron cluster with uniaxial anisotropy and hard axis along $z$-direction with strength of 0.9 mRy. The color bar is showing the $z$ component of normalized magnetization.}
    \label{fig:16}
\end{figure}

\begin{figure}[h]
    \centering
\includegraphics[width=0.44\textwidth, height=0.7\textwidth]{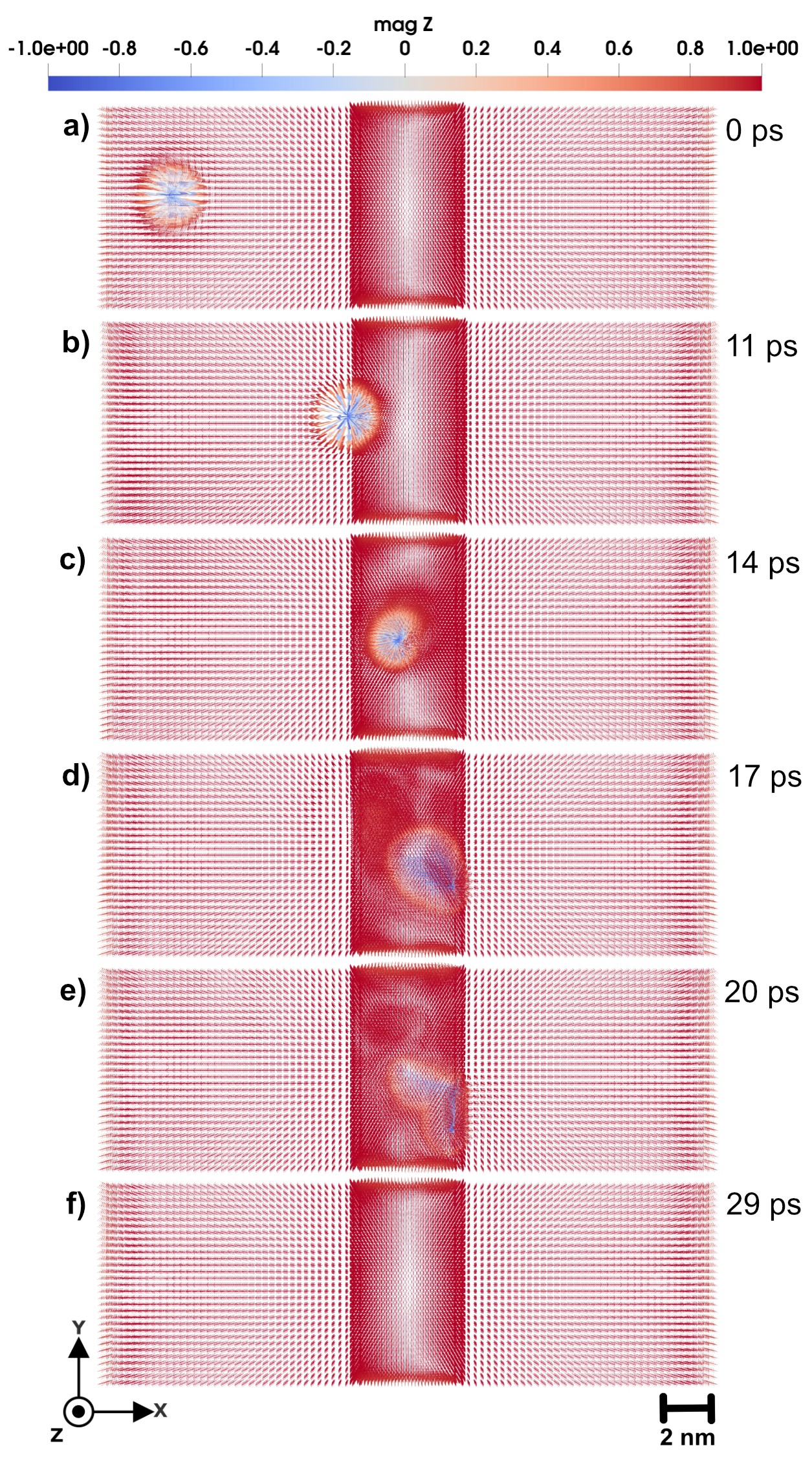}
    \caption{3D Skyrmion motion by applying STT of 30~m/s through the simulation box, with a defect region shaped as a tetrahedron cluster with uniaxial anisotropy and easy axis along $z$-direction with strength of 1.5 mRy. The color bar is showing the $z$ component of normalized magnetization.}
    \label{fig:17}
\end{figure}

At the highest defect anisotropy of 1.5 mRy, by applying STT with velocity $30~m/s$, the skyrmion exhibits  similar behavior to that observed at 0.9 mRy. When the anisotropy of the cluster is aligned along the easy-axis, the skyrmion disappears upon entering the tetrahedral region, as shown in Fig.~\ref{fig:17} and Supplementary Material, Video 15 \cite{Supplementary}.

\begin{figure}[h]
    \centering
 \includegraphics[width=0.44\textwidth, height=0.7\textwidth]{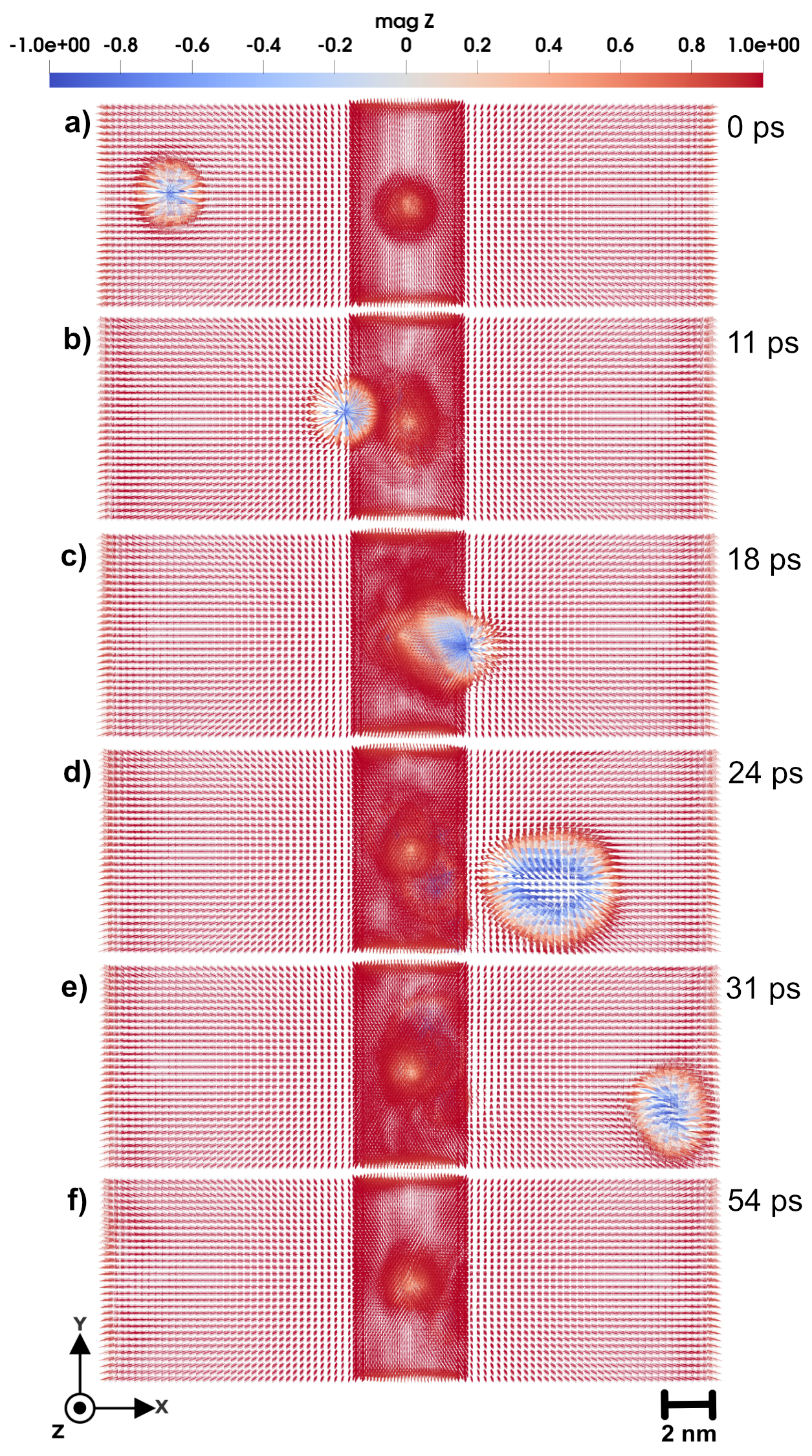}
    \caption{3D Skyrmion motion by applying a STT of 30~m/s through the simulation region, with a tetrahedron cluster with uniaxial anisotropy hard axis along lying in $x$-$y$ plane with strength of 1.5 mRy. The color bar is showing the $z$ component of normalized magnetization.}
    \label{fig:18}
\end{figure}

With the cluster anisotropy aligned with the hard axis, the skyrmion traverses the tetrahedral region, exhibiting a temporary expanding in size during the interaction, as illustrated in Fig.~\ref{fig:18}. This temporary size modulation is consistent with a breathing mode, a dynamic mode where the skyrmion radius oscillates in time due to perturbations in the local magnetic energy landscape \cite{Makhfudz2012, Kim2014, Onose2012, Mochizuki2012}. In this case, the abrupt change in anisotropy energy within the tetrahedron acts as a perturbative factor, inducing a nonlinear response of the skyrmion internal structure, leading to temporary expansion before relaxing back to its equilibrium size as shown in Fig.~\ref{fig:18} d)-f). The skyrmion gradually returns to its original size, indicating a recovery of its initial configuration once it exits the high-anisotropy area. The dynamics of the system is presented in Supplementary Material, Video 16 \cite{Supplementary}. This behavior suggests that engineered anisotropy regions can serve as control points for skyrmion deformation and energy storage, which are crucial for applications in skyrmion-based memory and logic devices where controlled skyrmion dynamics are essential. 

By decreasing the velocity of STT, the same trend remains as presented in Supplementary Material, Section IV, Figs.~S6-S9 and Videos 17-20 \cite{Supplementary}.

Therefore, in the cases where the anisotropy of the cluster is aligned along the hard axis of the micromagnetic region, and for varying anisotropy strengths of 0.11, 0.9, and 1.5 mRy, the skyrmion exhibits an increase in size after traversing the cluster and undergoes a breathing mode. In contrast, when the cluster anisotropy is oriented along the easy axis of the micromagnetic region, the skyrmion passes through the defect without any noticeable change in size or shape for low anisotropy values. However, at higher anisotropy strengths, the skyrmion is annihilated upon encountering the defect.

\section{CONCLUDING REMARKS}

In this work, we have developed and applied a fully 3D multiscale modeling approach to investigate the influence of defects on magnetization dynamics in hybrid micromagnetic-atomistic systems. By employing the $\mu$-ASD module, we explored the interplay between domain walls, skyrmions, and engineered local anisotropies in Fe-Ir thin film systems. Our simulations demonstrate that localized magnetic inhomogeneities, modeled as geometrically confined anisotropic clusters, exert a profound impact on both static and dynamic magnetic configurations. 

The double-slit structure enabled the observation of spin-wave interference fringes with wavelength modulations on the order of 20~\AA, in quantitative agreement with analytical predictions based on wave theory (Fig.~\ref{fig:4}) with a coefficient of determination $R^2 \approx 0.7$. This finding is analogous to quantum mechanical systems, reinforcing the wave nature of magnons in confined geometries. The results of this study contribute to the broader field of magnonics, where spin waves are used for information processing and wave-based computing. Understanding spin wave interference in structured media is crucial for developing next-generation magnonic logic devices, wave-based transistors, and nonvolatile spin wave memory, where information is encoded in the phase and amplitude of propagating magnons rather than in electron charge.
Additionally, domain wall dynamics in the presence of such slits revealed transient pinning and curvature effects, including enhanced post-slit acceleration explained through an effective energy-based model (Fig.~\ref{fig:4}). 

In the case of tetrahedral defect clusters with tunable anisotropy, we observed complex domain wall responses ranging from minor deformations and Barkhausen-like pinning (Fig.~\ref{fig:6}) at low anisotropy ($K_u \approx 0.11$ mRy) to complete topological transformations at higher values ($K_u \geq 0.9$ mRy). Notably, the formation of $90^\circ$-bent (Fig.~\ref{fig:8}, \ref{fig:10}) or tubular skyrmion-like textures (Fig.~\ref{fig:9}, \ref{fig:11}) was observed, illustrating how local anisotropy gradients can be utilized to create magnetic textures with non-trivial shape.  Furthermore, 3D skyrmion dynamics under spin-transfer torque exhibited sensitivity to defect-induced anisotropy. Soft easy-axis anisotropic regions ($K_u^{easy} \approx 0.11$ mRy) permitted smooth skyrmion motion (Fig.~\ref{fig:13}), while strong easy-axis anisotropy ($K_u^{easy} \geq 0.9$ mRy) led to topological breakdown which lead to skyrmion disappearing (Fig.~\ref{fig:15}, \ref{fig:17}), and hard-axis anisotropy caused temporary size modulation consistent with skyrmion breathing modes (Fig.~\ref{fig:14}, \ref{fig:16}, and \ref{fig:18}).

This study reinforces the concept of topological protection in magnetic skyrmions. Skyrmions displayed minimal deformation as they traversed impurity regions in our simulations. In contrast, domain walls, which lack a non-trivial topological character, are severely perturbed by defect structures, often becoming pinned, distorted, or even deformed. These observations align with the theoretical understanding that the non-zero topological charge of skyrmions imparts a degree of robustness against local perturbations, which is essential for their reliable function in spintronic applications where disorder and imperfections are inevitable.

These findings also demonstrate that localized anisotropic perturbations can reconfigure magnetic textures, modulate skyrmion topology, and control domain wall motion, in addition to serving as obstacles. Through defect engineering, it is possible to induce, control, and stabilize such behaviors for next-generation spintronic and magnonic devices. Further research should focus on validating the proposed mechanisms under realistic conditions, both computationally and experimentally. Experimental implementations involving pulsed microwave fields or spin-polarized currents can enable direct observations of spin-wave interference and skyrmion-guided magnon transport in curved geometries. As a result of such efforts, 3D defect-engineered magnetic systems will remain key components of future wave-based computing architectures that are energy-efficient.  By bridging atomistic precision with micromagnetic applicability, this work establishes a methodological and conceptual framework for exploring defect mediated phenomena in multiscale magnetic systems. Spin-based memory, logic, and signal processing technologies can be significantly impacted by anisotropy and topology as shown here, with significant effects on spin dynamics.

\vspace{2.0 em}

\begin{acknowledgments}
We acknowledge the support of the Knut and Alice Wallenberg Foundation (KAW - Scholar program and WISE - Wallenberg Initiative Materials Science). Support also acknowledged from eSSENCE, STandUPP, the European Research Council through the ERC Synergy Grant 854843-FASTCORR and the Swedish Research Council (VR). The computations were enabled by resources provided by the National Academic Infrastructure for Supercomputing in Sweden (NAISS), partially funded by the Swedish Research Council.
\end{acknowledgments}

\setcounter{section}{0}
\setcounter{figure}{0}
\setcounter{equation}{0}
\setcounter{table}{0}
\makeatletter 
\renewcommand{\thetable}{SI\@arabic\c@table}
\renewcommand{\figurename}{FIG.}
\renewcommand \thefigure{S\@arabic\c@figure}
\makeatother

\begin{widetext}
\begin{center}
\Large\textbf{Micromagnetic-atomistic hybrid modeling of defect-induced magnetization dynamics \\ ---Supplemental Material---}
\end{center}

\begin{center}
Nastaran Salehi$^1$, Olle Eriksson$^{1,2}$, Johan Hellsvik$^3$, and Manuel Pereiro$^1$\\
\it{$^1$Department of Physics and Astronomy, Uppsala University, 751 21 Uppsala, Sweden\\
$^2$WISE - Wallenberg Initiative Materials Science for Sustainability, Department of Physics and Astronomy, Uppsala University, SE-751 20 Uppsala, Sweden\\
$^3$PDC Center for High Performance Computing, KTH Royal Institute of Technology, SE-100 44 Stockholm, Sweden}

\end{center}

\section{Magnon dispersion relation}
\noindent
Starting with a Hamiltonian that describes the fundamental interactions of the Fe-Ir system given by:

\begin{equation}
	\mathcal{H}=-\frac{1}{2}\sum_{ij}\mathbf{s}_i  \mathcal{J}_{ij} \mathbf{s}_j=-\frac{1}{2}\sum_{ij}\left( J_{ij}\mathbf{s}_i \mathbf{s}_j+\mathbf{D}_{ij}\mathbf{s}_i\times s_j \right)
	\label{eq:1}
\end{equation}
where:
\begin{equation}
	\mathcal{J}_{ij}=\begin{pmatrix}
J_{ij} &  D_{ij}^z & -D_{ij}^y\\
 -D_{ij}^z& J_{ij} & D_{ij}^x \\
D_{ij}^y& -D_{ij}^x & J_{ij} \\ 
\end{pmatrix}
\label{eq:2}
\end{equation}
The exchange interaction between atom $i$ with atomic magnetic moment $s_i$ and atom $j$ with atomic magnetic moment $s_j$ is represented by $J_{ij}$ while the components of the Dzyaloshinskii-Moriya (DM) interaction are given by $D_{ij}^{\alpha}$.
The equation of motion for the atomic magnetic moments is given by:
\begin{equation}
	-i \hbar \frac{\mathrm{d} \mathbf{s}_k}{\mathrm{d}t}=\left[\mathcal{H},\mathbf{s}_k\right]=-\frac{1}{2}\left(\left[\sum_{ij}J_{ij} \mathbf{s}_i \mathbf{s}_j,\mathbf{s}_k \right]+\left[ \sum_{ij}\mathbf{D}_{ij} \mathbf{s}_i \times \mathbf{s}_j , \mathbf{s}_k\right] \right)
\label{eq:3}
\end{equation}

If we recast Eq.~\ref{eq:3} in components, then:

\begin{equation}
	-i \hbar \frac{\mathrm{d} s_{k\eta}}{\mathrm{d}t}=-\frac{1}{2}\left(\left[\sum_{ij}J_{ij} s_{i\alpha} s_j^\alpha,s_{k\eta} \right]+\left[ \sum_{ij}D^\alpha_{ij}\epsilon_{\alpha\beta\gamma} s_i^\beta s_j^\gamma, s_{k\eta}\right] \right)
    \label{eq:4}
\end{equation}

As notation, we use the latin letter for atomic positions while the greek letters are used for the components of the vector or tensor. By using the commutation relations of the spins:

\begin{equation}
 \begin{aligned}
	\left[s_{k\eta},s_{i\alpha}\right]&=-i g \mu_B \delta_{ki} \epsilon_{\eta\alpha\gamma}s_k^\gamma \\
	\left[s_{k\eta},s_j^\alpha\right]&=-i g \mu_B \delta_{kj} \epsilon_{\eta\gamma}^\alpha s_k^\gamma,
\end{aligned}
\label{eq:5}
\end{equation}

and after some algebra, the different terms of the equation of motion are given by:
\begin{equation}
\begin{aligned}
	\left[\sum_{ij}J_{ij} \mathbf{s}_i \mathbf{s}_j,\mathbf{s}_k \right]&=2 i g \mu_B\sum_{j\neq k} J_{kj} \epsilon_{\eta\alpha\gamma} s_k^\gamma s_j^\alpha \\
	\left[ \sum_{ij}D^\alpha_{ij}\epsilon_{\alpha\beta\gamma} s_i^\beta s_j^\gamma, s_{k\eta}\right]&=2 i g \mu_B\sum_{j\neq k} D_{kj}^\alpha \epsilon_{\alpha\beta\gamma} \epsilon_{\eta\chi}^\gamma s_k^\chi  s_j^\beta,
\end{aligned}	
\label{eq:6}
\end{equation}

Assuming the atomic magnetic moment of the system is in the ground state with the moments pointing almost along the z direction, i.e., $\mathbf{s}_i=s_i \mathbf{\hat{e}}_3$ with $\mathbf{\hat{e}}_3$ being a orthogonal unit vector as indicated in Fig.~\ref{fig:s1}. We perturb the ground-state by using a spin-spiral excitation, and so the magnetic moment vector can be recasted in spherical coordinates like:

\begin{equation}
	\mathbf{s}_i=s_i\left(\sin\theta_i \cos\phi_i \mathbf{\hat{e}}_1+\sin\theta_i\sin(c\phi_i) \mathbf{\hat{e}}_2 + \cos(\theta_i) \mathbf{\hat{e}}_3\right)
    \label{eq:7}
\end{equation}
where $\mathbf{\hat{e}}_1$ and $\mathbf{\hat{e}}_2$ are orthogonal unit vectors as shown in Fig.~\ref{fig:s1} and $c=\pm 1$ represents the chirality of the spin waves.

\begin{figure}[htp]
\centering
\includegraphics[width=0.3\linewidth]{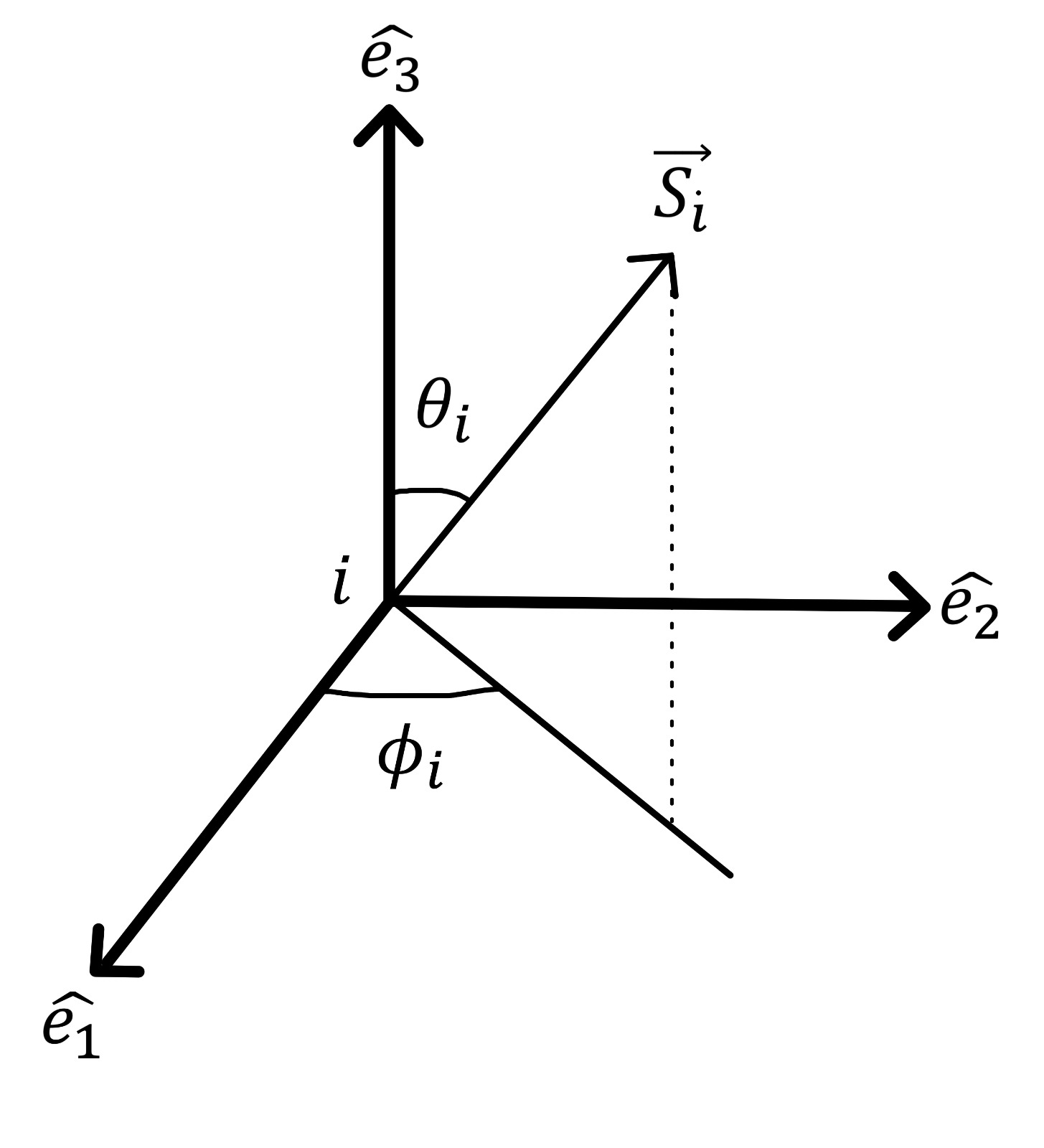}  
\caption{(Color online) Representation of the components of the atomic magnetic moment of atom $i$ in spherical and Cartesian coordinates.}
\label{fig:s1}
\end{figure}
Keeping in mind that we are disregarding Stoner excitations, in which the length of the magnetic moment is considered constant as time evolves, we obtain, after some algebra, the equation of motion:

\begin{equation}
\begin{aligned}
 \frac{\mathrm{d} s_i}{\mathrm{d}t}&=0 \\
 c\sin\theta_i  \frac{\mathrm{d} \mathbf{\phi}_i}{\mathrm{d}t}&= \frac{g \mu_B}{\hbar} \sum_{j\neq i } s_j[J_{ij} (-\cos\theta_i \sin\theta_j \cos(\phi_i-\phi_j)+\sin\theta_i\cos\theta_j)-(\mathbf{D}_{ij}\mathbf{\hat{e}}_1)\sin\theta_i\sin\theta_j\sin(c\phi_j) \\
 &-(\mathbf{D}_{ij}\mathbf{\hat{e}}_1)\sin(c\phi_i)\cos\theta_i\cos\theta_j 
 + (\mathbf{D}_{ij}\mathbf{\hat{e}}_2)\sin(\theta_i)\sin\theta_j\cos\phi_j \\
 &+ (\mathbf{D}_{ij}\mathbf{\hat{e}}_2)\cos(\phi_i)\cos\theta_i\cos\theta_j+ (\mathbf{D}_{ij}\mathbf{\hat{e}}_3)\cos(\theta_i)\sin\theta_j\sin(c(\phi_i-\phi_j)] \\
  \frac{\mathrm{d} \mathbf{\phi}_i}{\mathrm{d}t}&= \frac{-g \mu_B}{\hbar} \sum_{j\neq i } s_j[J_{ij} \sin\theta_j \sin(c(\phi_i-\phi_j)) -(\mathbf{D}_{ij}\mathbf{\hat{e}}_1)\cos\phi_i\cos\theta_j -(\mathbf{D}_{ij}\mathbf{\hat{e}}_2)\sin(c\phi_i)\cos\theta_j \\
 &+ (\mathbf{D}_{ij}\mathbf{\hat{e}}_3)\sin\theta_j\cos(\phi_i-\phi_j)] \\
 \label{eq:8}
\end{aligned}
\end{equation}

Far from the point sources that produce the magnetic perturbation, we can assume that $\theta_i$ is constant and consequently, $\frac{\mathrm{d} \mathbf{\theta}_i}{\mathrm{d}t}=0$. Then, in Eq.~\ref{eq:8}, by summing up the second equation with the third one multiplied by $ i c \cos\theta_i \frac{\mathrm{d} \mathbf{\theta}_i}{\mathrm{d}t}$ and by time-averaging it, we end up in the following expression:

\begin{equation}
	\theta_i\frac{\mathrm{d}(c\phi_i)}{\mathrm{d}t}=\frac{g\mu_B}{\hbar}\sum_{j\neq i}s_j[-J_{ij}(\theta_j e^{i (\phi_i-\phi_j)}-\theta_i)-i c(\mathbf{D}_{ij} \mathbf{\hat{e}}_3)\theta_j e^{i (\phi_i-\phi_j)}]
	\label{eq:9}
\end{equation}

The Fourier transform of Eq.~\ref{eq:9} is:

\begin{equation}
	\theta_\mu \omega^c=\frac{g \mu_B}{\hbar} \sum_{[\nu+\mathbf{R}]\neq \mu} s_{[\nu+\mathbf{R}]}[-J_{[\nu+\mathbf{R}]}( \theta_{[\nu+\mathbf{R}]} e^{i(q(\tau_\mu-\tau_\nu-\mathbf{R})+\phi_\mu-\phi_\nu)}-\theta_\mu)-ic(\mathbf{D}_{\mu[\nu+\mathbf{R}]} \mathbf{\hat{e}}_3) \theta_{[\nu+\mathbf{R}]} e^{i(q(\tau_\mu-\tau_\nu-\mathbf{R})+\phi_\mu-\phi_\nu)}]
    \label{eq:10}
\end{equation}
where the following notation has been used: $\mathbf{R}_i=\tau_\mu$ (atomic position of atom $i$), $\mathbf{R}_j=\mathbf{R}+\tau_\mu$ (atomic position of atom $j$ and $\mathbf{R}$ is the lattice displacement), $\phi_i=\omega t+ \mathbf{q} \tau_\mu+\phi_\mu$ (phase angle of atom $i$), $\phi_j=\omega t+ \mathbf{q} \tau_\nu+\phi_\nu$ (phase angle of atom $j$), $\phi_i-\phi_j=\mathrm{q}(\tau_\mu-\mathbf{R})+\phi_\mu-\phi_\nu$ (phase difference between atom $i$ and $j$, respectively), $i\rightarrow\mu$, $j\rightarrow[\nu+\mathbf{R}]$. Moreover, we also use that $\frac{\mathrm{d}(c\phi_i)}{\mathrm{d}t}=\omega_i^c=\mathrm{constant}=\omega^c \; \forall i$. By following this change of notation, from now on, the greek letters $\mu$ and $\nu$ will refer also to atomic positions rather than vector coordinates. In Eq.~\ref{eq:9}, we have also Fourier transformed the coupling interactions as given by:
\begin{equation}
\begin{aligned}
J_{\mu\nu}(\mathrm{\mathbf{q}})=\delta_{\mu\nu}J_{\mu\mu}-\sum_\mathbf{R} J_{\mu[\nu+\mathbf{R}]} e^{i(q(\tau_\mu-\tau_\nu-\mathbf{R})}\\
\mathbf{D}_{\mu\nu }=\delta_{\mu\nu}\mathbf{D}_{\mu\mu}-\sum_	\mathbf{R} \mathbf{D}_{\mu[\nu+\mathbf{R}]} e^{i(q(\tau_\mu-\tau_\nu-\mathbf{R})}
\end{aligned}
\label{eq:11}
\end{equation}
In the ansatz of the current formulation, it follows that:
\begin{equation}
	s_{[\nu+\mathbf{R}]}=s_\nu; \; \theta_{[\nu+\mathbf{R}]}=\theta_\nu 
    \label{eq:12}
\end{equation}
Then, Eq.~\ref{eq:9} can be recasted as:
\begin{equation}
\tilde{\theta}_\mu \hbar \omega^c=g\mu_B\sum_{\nu\neq \mu}\sqrt{s_\mu}\sqrt{s_\nu } \tilde{\theta}_\nu[J_{\mu\nu}(\mathbf{q})-\delta_{\mu\nu}\sum_\lambda \frac{s_\lambda}{s_\mu}J_{\mu\lambda}(\mathbf{q}=0)+i c (\mathbf{D}_{\mu\nu}(\mathbf{q}) \mathbf{\hat{e}}_3)]
\label{eq:13}
\end{equation}
where we defined $\tilde{\theta}_\mu=\sqrt{s_\mu}\theta_\mu e^{-i \phi_\mu} $. If we define
\begin{equation}
	\tilde{T}_{\mu\nu}(\mathbf{q})=J_{\mu\nu}(\mathbf{q})-\delta_{\mu\nu}\sum_\lambda\frac{s_\lambda}{s_\mu}J_{\mu\lambda}(\mathbf{q}=0)+ic(\mathbf{D}_{\mu\nu}(\mathbf{q}) \mathbf{\hat{e}}_3) 
    \label{eq:14}
\end{equation}
the final from of Eq.~\ref{eq:13} can be given by:
\begin{equation}
	\tilde{\theta}_\mu \hbar \omega^c=g\mu_B\sum_{\nu\neq\mu}\sqrt{s_\mu}\sqrt{s_\nu}\tilde{T}_{\mu\nu}(\mathbf{q}) \tilde{\theta}_\nu 
	\label{eq:15}
\end{equation}
By solving the eigenvalue problem described by Eq.~\ref{eq:15}, it is only required to diagonalise the following tensor:
\begin{equation}
	\mathcal{A}_{\mu\nu}=\frac{g\mu_B}{\hbar}\sum_{\nu\neq\mu}\sqrt{s_\mu}\sqrt{s_\nu} \tilde{T}_{\mu\nu}(\mathbf{q}) 
    \label{eq:16}
\end{equation}
The eigenvalues of $\mathcal{A}_{\mu\nu}$ matrix provide the adiabatic magnon spectra as a function of the wave vector in the reciprocal space.

Assuming that the magnetic texture of the material have a small angle with respect to the $\mathbf{\hat{e}}_3$ direction, we can assume that $\tilde{\theta}_\mu\approx\tilde{\theta}_\nu $. Moreover, for simplicity without loss of generality we assume $s_\mu=s_\nu=s$. Then, Eq.~\ref{eq:15} can be recasted as:
\begin{equation}
 \hbar \omega^c=g\mu_B s \sum_{\nu\neq\mu}\tilde{T}_{\mu\nu}(\mathbf{q}) 
 \label{eq:17}
\end{equation}

For small $\mathbf{q}$-vector and considering that $\mu\neq\nu$, the exchange part of the magnon energies can be reduced to:
\begin{equation}
  \Re(J_{\mu\nu}(\mathbf{q}))\simeq \frac{1}{2} \sum_\mathbf{R} J_{\mu [\nu+\mathbf{R}]} \mathbf{q}^2 (\tau_\mu-\tau_\nu-\mathbf{R})^2 
  \label{eq:18}
\end{equation}

Considering that the exchange interaction is relevant for only first neighbouring atoms and all atoms present the same interaction (J):
\begin{equation}
	\Re(J_{\mu\nu}(\mathbf{q}))\simeq \frac{J}{2}z a^2 q^2
    \label{eq:19}
\end{equation}
where z is coordination number and a is the lattice constant and the dispersion relation for the exchange part of the hamiltonian is given by:
\begin{equation}
	\hbar \omega=\frac{1}{2}g\mu_B s J (z a)^2 q^2
    \label{eq:20}
\end{equation}
where $D_{ex}=\frac{1}{2}g\mu_B s J (z a)^2$ represents the spin-wave stiffness. 

\begin{figure}[htp]
\centering
\includegraphics[width=0.5\linewidth]{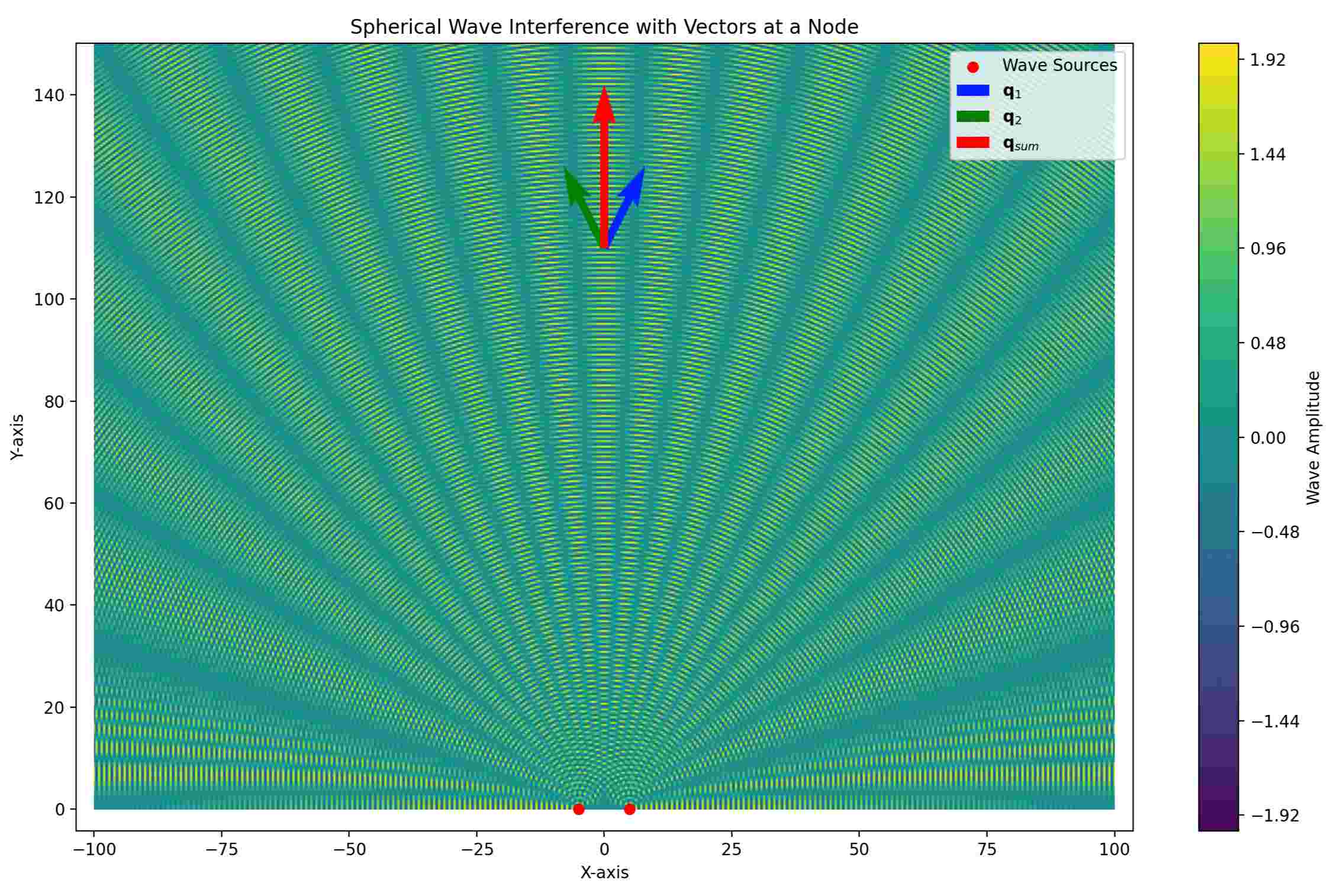}  
\caption{(Color online) Schematic representation of a spherical wave interference pattern provided by 2 point wave sources shown in red. Far from the sources, in the central region the wave vector $\vec{k}$ can be approximated to be aligned along the y direction as shown by the red arrow which is the sum of the individual wave vectors (in blue and green color) produced by the red sources, respectively.}
\label{fig:s2}
\end{figure}

For the DM interaction with small $\mathbf{q}$, we have:
 
\begin{equation}
\Re({\mathbf{D}_{\mu\nu}})=-\sum_\mathbf{R}c \mathbf{D}_{\mu[\nu+\mathbf{R}]}^z\left[\mathbf{q}(\tau_\mu-\tau_\nu-\mathbf{R})\right]
\label{eq:21}
\end{equation}
Assuming as in the case of the exchange interaction, that DM interaction only is relevant for the first-neighbours and it has the same strength (D) as it is the case for the current system considered in this article, 
 \begin{equation}
 	\Re({\mathbf{D}_{\mu\nu}})=-c D q \sum_\mathbf{R}\mathbf{D}_{\mu[\nu+\mathbf{R}]}^z |\tau_\mu-\tau_\nu-\mathbf{R}| \cos\theta(\mathbf{q},\mathbf{R})
    \label{eq:22}
 \end{equation}
For the simple cubic case, $\mathbf{R}=\{a(\pm 1,0,0), a(0,\pm 1,0),a(0,0,\pm 1)\}$, the DM term can be finally recasted in:
\begin{equation}
	\Re({\mathbf{D}_{\mu\nu}})=-2 c D a (q^x+q^y+q^z) 
    \label{eq:23}
\end{equation}
As shown in Fig.~\ref{fig:s2}, for a spherical wave pattern provided by 2 point wave sources far from the sources, the total wave vector in the central region almost only the  x component survives, so that we can approximate the DM term by:
\begin{equation}
	\Re({\mathbf{D}_{\mu\nu}})\simeq-2 c D a q 
    \label{eq:24}
\end{equation} 

In conclusion the dispersion relation for the hamiltonian given in Eq.~\ref{eq:1} can be recasted in the following expression:
\begin{equation}
	\hbar\omega^c=D_{ex} q^2 - c D_{dm} q
	\label{eq:25}
\end{equation}
where $D_{dm}=g \mu_B s D z^2 a$.

Finally, the wave vector can be given by:
\begin{equation}
	q=\frac{c D_{dm}\pm\sqrt{D_{dm}^2\mp4D_{ex}\hbar \omega }}{2D_{ex}}
    \label{eq:26}
\end{equation}

\section{Model on the diffraction pattern of the macromagnetic spin waves}

\begin{figure}[htp]
\centering
\includegraphics[width=0.5\linewidth]{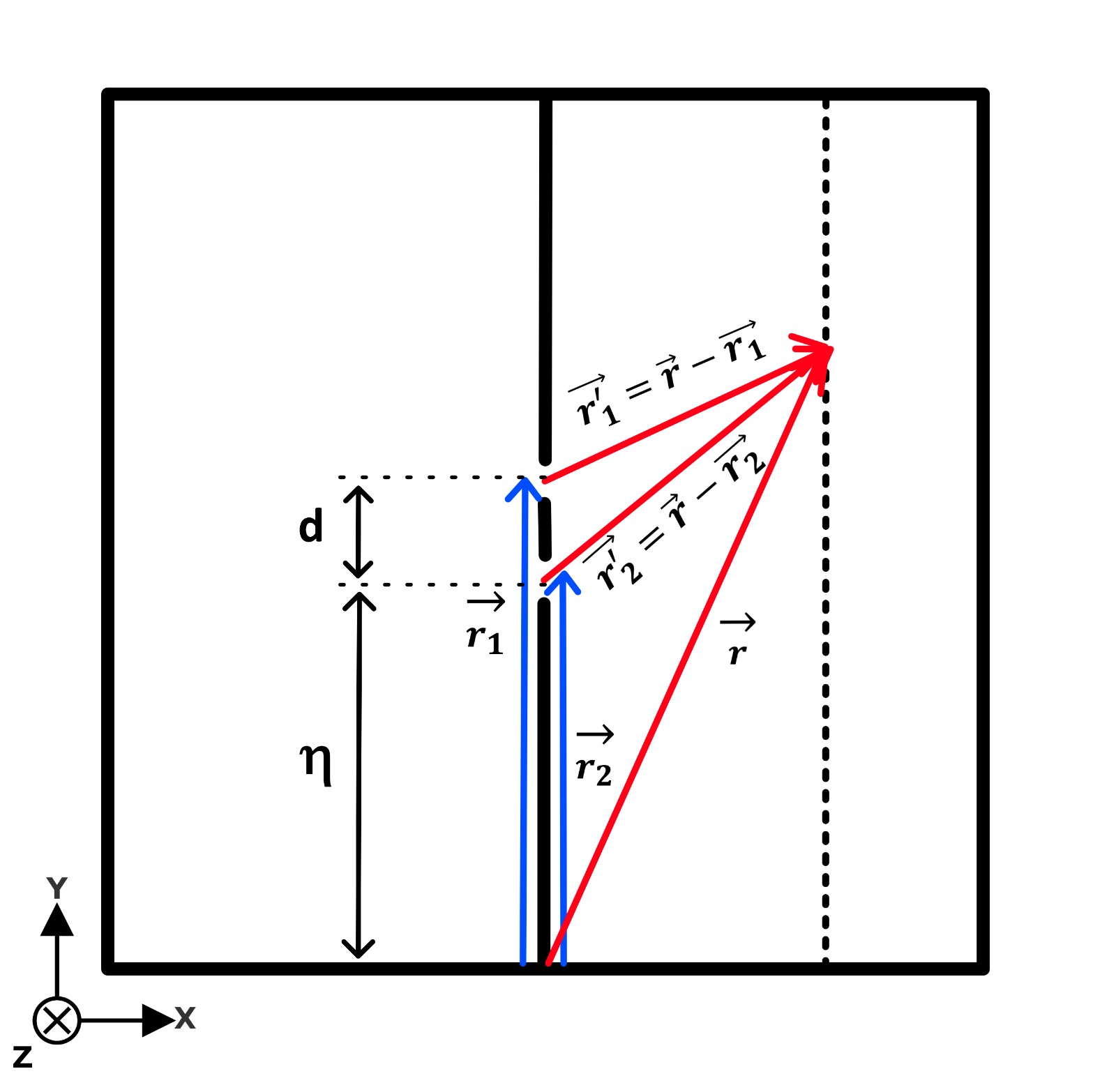}  
\caption{(Color online) Schematic representation of the different vectors and parameter used to describe the propagation of the spin waves. The symbol d represents the distance between the slits while $\eta$ accounts for the distance between the reference frame and the first slit. The vectors are described by the red and blue arrows.}
\label{fig:s3}
\end{figure}

A spherical wave, generated by two point sources whose intensity decays exponentially, can be described mathematically using the following expression:
\begin{equation}
	\mathbf{W}_{ext}=W_1 \frac{e^{i(q|\mathbf{r}-\mathbf{r}_1|-\omega t)}}{|\mathbf{r}-\mathbf{r}_1|} e^{|\mathbf{r}-\mathbf{r}_1|\xi}+W_2 \frac{e^{i(q|\mathbf{r}-\mathbf{r}_2|-\omega t)}}{|\mathbf{r}-\mathbf{r}_2|}e^{|\mathbf{r}-\mathbf{r}_2|\xi}
    \label{eq:27}
\end{equation}
where $\xi$ is the attenuation coefficient induced by the medium where the spin waves propagate and $W_1$ and $W_2$ are the amplitudes of the spherical waves created in slits 1 and 2, repectively. The definition of the vectors and notation of the parameters used on this section is shown in Fig.~\ref{fig:s3}. The intensity of the wave function is obtaining by multiplying the wave function by the complex conjugate, so that:
\begin{equation}
	|W_{ext}|^2=W_0^2\left[\frac{e^{2|\mathbf{r}-\mathbf{r}_1|\xi}}{|\mathbf{r}-\mathbf{r}_1|^2} +\frac{e^{2|\mathbf{r}-\mathbf{r}_2|\xi}}{|\mathbf{r}-\mathbf{r}_2|^2} +\frac{2\cos(q(|\mathbf{r}-\mathbf{r}_1|-|\mathbf{r}-\mathbf{r}_2|))e^{(|\mathbf{r}-\mathbf{r}_1|+|\mathbf{r}-\mathbf{r}_2|)\xi}}{|\mathbf{r}-\mathbf{r}_1| |\mathbf{r}-\mathbf{r}_2| } \right]
    \label{eq:28}
\end{equation}
where, for simplicity, we have assumed that $W_1=W_2=W_0$. Following the setup and the reference frame shown in Fig.~\ref{fig:s3}, the coordinates of the vectors are given by:  $\mathbf{r}_1=(0,d+\eta)$, $\mathbf{r}_2=(0,\eta)$, $\mathbf{r}=(x,y)$, $\mathbf{r}-\mathbf{r}_1=(x, y-(d+\eta))$ and $\mathbf{r}-\mathbf{r}_2=(x, y-\eta)$. The final expression of the wave amplitude in the reference frame pictured in Fig.~\ref{fig:s3} is:
\begin{equation}
\begin{aligned}
	|W_{ext}|^2&=W_0^2\left[\frac{e^{2\sqrt{x^2+(y-(d+\eta))^2}\xi}}{x^2+(y-(d+\eta))^2} +\frac{e^{2\sqrt{x^2+(y-\eta)^2}\xi}}{x^2+(y-\eta)^2} \right. \\ & \left.  +\frac{2\cos(q(\sqrt{x^2+(y-(d+\eta))^2}-\sqrt{x^2+(y-\eta)^2}))e^{(\sqrt{x^2+(y-(d+\eta))^2} +\sqrt{x^2+(y-\eta)^2})\xi}}{\sqrt{x^2+(y-(d+\eta))^2} \sqrt{x^2+(y-\eta)^2} } \right]
\end{aligned}
\label{eq:29}
\end{equation}

Finally, for small angle deviations we can approximate the average angle deviation squared by:

\begin{equation}
\langle \theta \rangle^2\simeq \langle\frac{\gamma W_{ext}}{\omega} \rangle^2 + \sigma=\left(\frac{\gamma}{\omega}\right)^2\langle W_{ext}\rangle^2 +  \sigma
\label{eq:30}
\end{equation}
where $\sigma$ represents background intensity, potentially arising from microwave field-driven energy input. The attenuation coefficient $\xi$ is the imaginary part of the frequency of the spin wave and it is related to the Gilbert damping $\alpha$ by:

\begin{equation}
	\xi=\frac{\alpha \omega}{v_g}
    \label{eq:31}
\end{equation}
where $v_g$ is the group velocity and it is defined as $\mathbf{v}_g=\frac{\partial\omega}{\partial \mathbf{q}}$. By using Eq.~\ref{eq:25}, the attenuation coefficient can finally be derived as:
\begin{equation}
	\xi^c=\frac{\alpha\omega}{2 \frac{D_{ex}}{\hbar}q-c \frac{D_{dm}}{\hbar}}
    \label{eq:32}
\end{equation}

The parameters used in the spin wave model to predict the form of the diffraction pattern are collected in Table~\ref{table:1}.

\begin{table}[h!]
\centering
\caption{Table of parameters used in the spin wave model to describe the diffraction pattern.}
\begin{tabularx}{400pt}{@{\extracolsep{\fill}}lrr@{} }
\hline
\hline
Quantity & Magnitude & unit\\
\hline
$\eta$  & 80 & \AA  \\
d & 20 &\AA   \\
x & 82 &\AA   \\
$\xi$ & -0.561918 &\AA$^{-1}$   \\
$\alpha$ & 0.1 &  \\
f & 1  & $10^{12}$ Hz  \\
$\gamma$ & 176085.81 & $10^6$ $\frac{Hz}{T}$  \\
z & 6&  \\
q & -1.52116&\AA$^{-1}$\\
D &   0.28774338& $10^{-21}$ J \\
J &  0.91336725   & $10^{-21}$ J  \\
$\sigma$ & $0.0254$ & degree$^2$ \\
$W_0$ & $44800$& degree $\cdot$ T\\
$s$  & $2.23$ &$\mu_B$ \\
\hline
\hline
\end{tabularx}
\label{table:1}
\end{table}

\section{Theoretical one-dimensional (1D) model for domain wall acceleration after geometric constriction by the slit}

The double slit imposes a geometric constraint, forcing the DW to deform. This deformation stores potential energy, modelled by the term $V(q)$. We assume $V(q)$ has a maximum value $V_{0}$ at the slit position $q^\prime$, representing the elastic energy stored due to compression against exchange and anisotropy forces. The potential can be mathematically modeled by a smooth Dirac delta in spatial dimension by using the following expression:

\begin{equation}
	V(q-q\prime)=V_0 \; e^{-\frac{q-q\prime}{a}}
    \label{eq:33}
\end{equation}
where $a$ is the width of the potential.

\begin{figure}[h]
    \centering
\includegraphics[width=0.7\textwidth, height=0.7\textwidth]{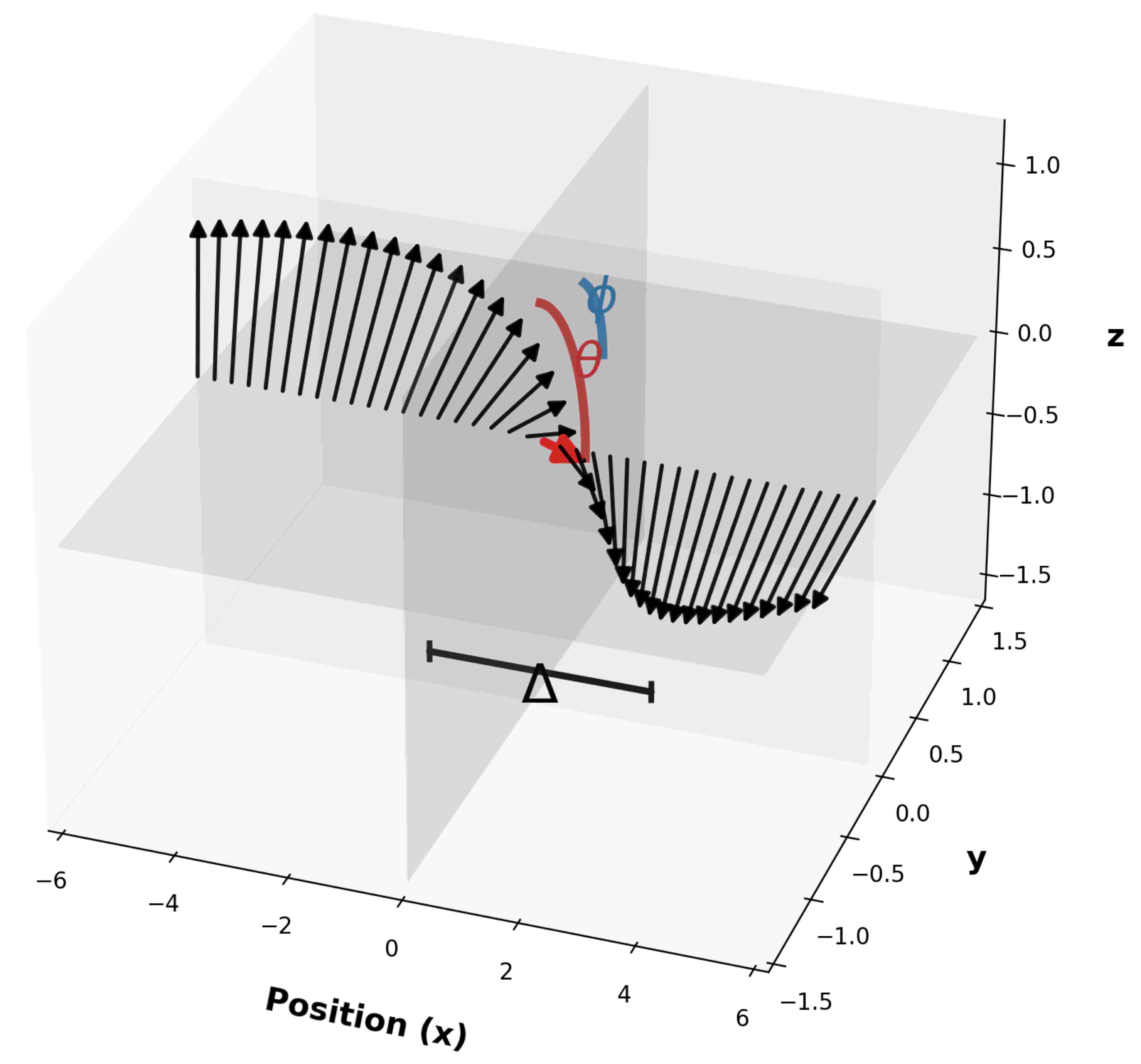}
    \caption{1D domain wall indicating the angles $\theta$, $\phi$ and the domain wall width $\Delta$.}
    \label{fig:s4}
\end{figure}

We now consider a 1D model of the DW dynamics subject to the action of a field H perpendicular to the $x$-axis and aligned along the $z$-direction using the reference frame depicted in Fig.~\ref{fig:s4}. The magnetization is represented in spherical coordinates as indicated in Fig.~\ref{fig:s5}. The one-dimensional domain wall is fully characterized by three parameters: $q(x,t)$ specifies the wall center position along the $x$-axis, $\Delta(x,t)$ determines the domain wall width, and $\phi(x,t)$ defines the magnetization rotation angle. The resulting magnetization profile is described everywhere by the function:

\begin{eqnarray}
    \theta(x, t) &=& 2 \tan^{-1} \left[ \exp\left( \frac{x - q(t)}{\Delta(t)} \right) \right]  \label{eq:34} \\
    \varphi(x, t) &=& \phi(t)
    \label{eq:35}
\end{eqnarray} 
where $\varphi(x, t)$ is the angle between the $y$-axis and the plane formed by the magnetization vector and the projection of the magnetization over the $y$-$z$ plane as shown in Fig.~\ref{fig:s5}. For simplicity, it is assumed to be independent of the position and only can change in time.

\begin{figure}[h]
    \centering
\includegraphics[width=0.5\textwidth]{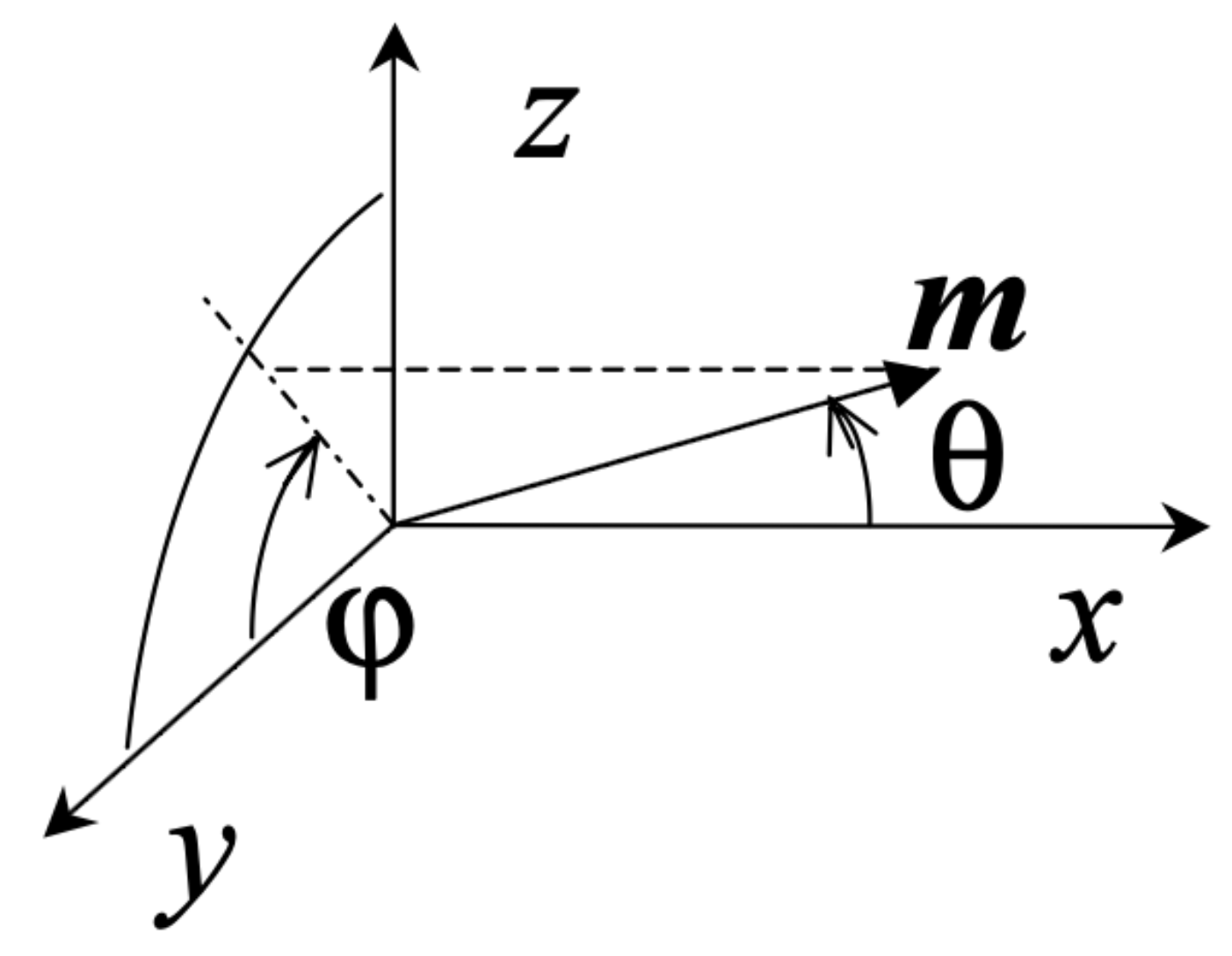}
    \caption{Reference frame used to define the angles $\theta$ and $\varphi$ of the magnetization in spherical coordinates.}
    \label{fig:s5}
\end{figure}

In order to obtain the equations of motion of the magnetization, and in particular, the velocity of the domain wall, we will start from the Lagrangian form of the micromagnetic dynamic equations:

\begin{equation}
    \frac{\partial L}{\partial \zeta} - \frac{\rm{d}}{\rm{d}t} \left( \frac{\partial L}{\partial \dot{\zeta}} \right) + \frac{\partial F}{\partial \dot{\zeta}} = 0
    \label{eq:36}
\end{equation}
 where $\zeta$ is the generalized coordinate and can be represented by $q$, $\Delta$ or $\phi$. The Lagrange function is $ \mathcal{L} = \int_0^\infty \left( \mathcal{E} + \frac{M_s}{\gamma} \dot{\phi} \cos\theta \right) dx - V(q-q^\prime)$ and the dissipation energy is $\mathcal{F} = \left(\frac{\alpha M_s}{2\gamma}\right) \int_0^\infty \left( \dot{\theta}^2 + \dot{\phi}^2 \sin^2\theta \right) dx$ \cite{Thiaville2006}. Moreover, $\mathcal{E}$ is the micromagnetic density energy, $\gamma$ is the gyromagnetic ratio, $\alpha$ is the Gilbert damping and $M_s$ is the saturation magnetization. The micromagnetic energy density for a 1D domain wall model, considering contributions from exchange ($A$), Dzyaloshinskii-Moriya interaction ($D$), uniaxial anisotropy ($K_0$), second-order uniaxial transverse anisotropy ($K$), and an external magnetic field ($H$), is given by:
 
\begin{equation}
   \mathcal{E} = A \left(\frac{\partial{\theta}}{\partial{x}}\right)^2 +\frac{D}{\Delta} \sin\phi \sin\theta+ K_0 \sin^2\theta + K \sin^2\theta \sin^2\phi - \mu_0 M_s H \cos\theta
   \label{eq:37}
\end{equation}
 where $\mu_0$ is the vacuum magnetic permeability. Thus, the space-integrated Lagrangian $\mathcal{L}$ can be recasted in the following form:
 \begin{equation}
 	\mathcal{L}=\frac{2A}{\Delta}+\frac{\pi D}{2\Delta} \sin\phi+2\Delta \left(K_0+K\sin^2\phi\right)-2\mu_0 M_s H q+\frac{2\mu_0 M_s q}{\gamma} \dot{\phi}-V_0 e^{-\left(\frac{q-q^\prime}{a}\right)^2}
    \label{eq:38}
 \end{equation}
 
 Similarly, the space-integrated dissipation function $\mathcal{F}$ takes the form:
 \begin{equation}
 	\mathcal{F}=\frac{\alpha \mu_0 M_s}{\gamma} \left[\Delta \dot{\phi}^2+\frac{\dot{q}^2}{\Delta}+\frac{\dot{q}\dot{\Delta}}{\Delta} \ln2+\frac{\pi^2}{24} \left(\frac{\dot{\Delta}}{\Delta}\right)^2\right]
    \label{eq:39}
 \end{equation}
 
 Since we are just interested in the velocity of the domain wall, by setting $\zeta=\phi$ and $\zeta=q$ in the Euler-Lagrange equations (Eq.~\ref{eq:36}), we end up with the following equations of motion:
 \begin{eqnarray}
 	\frac{\dot{q}}{\Delta}-\alpha\dot{\phi}=\frac{\pi D \gamma}{4 \Delta^2 \mu_0 M_s} \cos\phi+\gamma H_k \sin(2\phi)
 	\label{eq:40} \\
 	-\gamma H + \dot{\phi}+\eta e^{-\left(\frac{q-q^\prime}{a}\right)^2} (q-q^\prime)+\alpha \left(\frac{\dot{q}}{\Delta}+\frac{\dot{\Delta}}{\Delta}\ln \sqrt{2}\right)=0
 	\label{eq:41}
 \end{eqnarray}
 where $H_k=\frac{K}{\mu_0 M_s}$ is the uniaxial anisotropy field and $\eta=\frac{\gamma V_0}{\mu_0 M_s a^2}$. Eliminating $\dot{\phi}$ from Eqs.~\ref{eq:40}-\ref{eq:41}, the velocity of the domain wall ($\dot{q}$) is finally given by:
 \begin{align}
    \dot{q} &= \frac{1}{1+\alpha^2} \left( \alpha \gamma \Delta H - \alpha \Delta \eta e^{-\left(\frac{q-q'}{a}\right)^2} (q-q') - \alpha^2 \dot{\Delta} \ln\sqrt{2} \right. \nonumber \\
    &\quad \left. + \frac{\pi D \gamma}{4\Delta \mu_0 M_s} \cos\phi + \gamma \Delta H_k \sin(2\phi) \right)
    \label{eq:42}
\end{align}

The chirality induced by the Dzyaloshinskii-Moriya interaction (DMI) affects the domain wall velocity, with the direction of the effect determined by the sign of the DMI constant, $D$. Specifically, positive $D$ tends to increase velocity, while negative $D$ tends to decrease it. However, focusing on the interaction with the slit potential, the primary mechanism for velocity increase upon traversing the slit involves the second and third terms of Eq.~\ref{eq:42}. The second term, originating from the slit potential gradient, initially accelerates the domain wall as it approaches the slit ($q-q^\prime < 0$), thereby yielding a positive contribution to the driving force. Although this acceleration is followed by a deceleration after the slit ($q-q^\prime > 0$, yielding a negative contribution), the initial acceleration phase critically modifies the domain wall structure. It induces an abrupt decrease in the domain wall width ($\Delta$), resulting in a negative time derivative ($\dot{\Delta}<0$). Consequently, the third term becomes positive, providing an additional driving force that significantly increases the domain wall velocity after it exits the slit. Indeed, simulations confirm this mechanism, showing that the velocity upon exiting the slit can be up to twice its value prior to encountering the slit.

\newpage

\section{Figures}
\begin{figure}[H]
    \centering
\includegraphics[width=0.44\textwidth, height=0.7\textwidth]{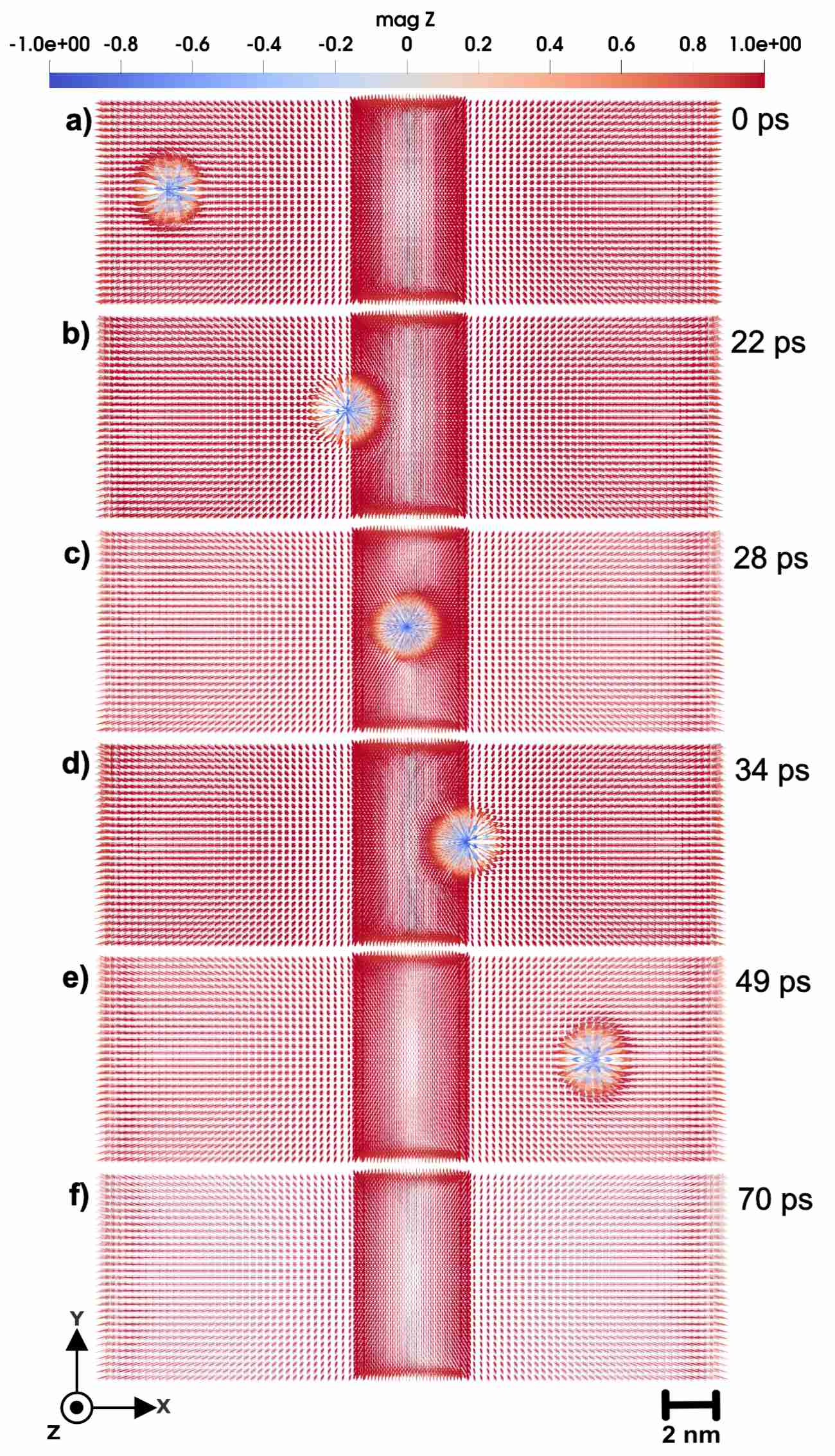}
    \caption{3D Skyrmion motion by applying STT of 15~m/s through the simulation cell, with a defect region shaped as a tetrahedron cluster with uniaxial anisotropy and hard axis along $z$-direction with strength of 0.11 mRy. The color bar is showing the $z$ component of normalized magnetization. The dynamics of the system is presented in Supplemental Material, Video 17 \cite{Supplementary}}
    \label{fig:s6}
\end{figure}

\begin{figure}[H]
    \centering
\includegraphics[width=0.44\textwidth, height=0.7\textwidth]{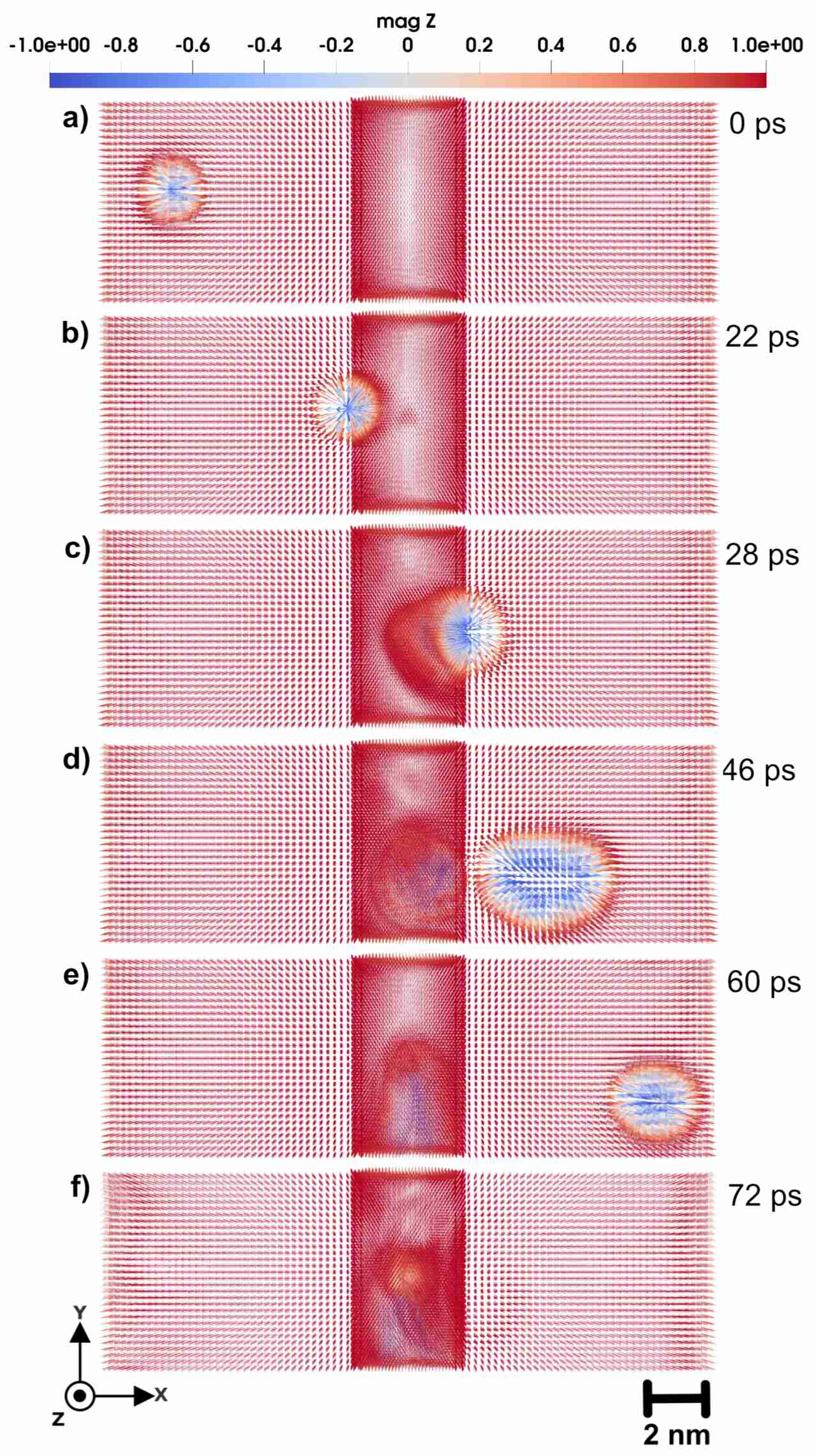}
    \caption{3D Skyrmion motion by applying STT of 15~m/s through the simulation box, with a defect region shaped as a tetrahedron cluster with uniaxial anisotropy and easy axis along $z$-direction with strength of 0.11 mRy. The color bar is showing the $z$ component of normalized magnetization. The dynamics of the system is presented in Supplemental Material, Video 18 \cite{Supplementary}}
    \label{fig:s7}
\end{figure}

\begin{figure}[H]
    \centering
\includegraphics[width=0.44\textwidth, height=0.7\textwidth]{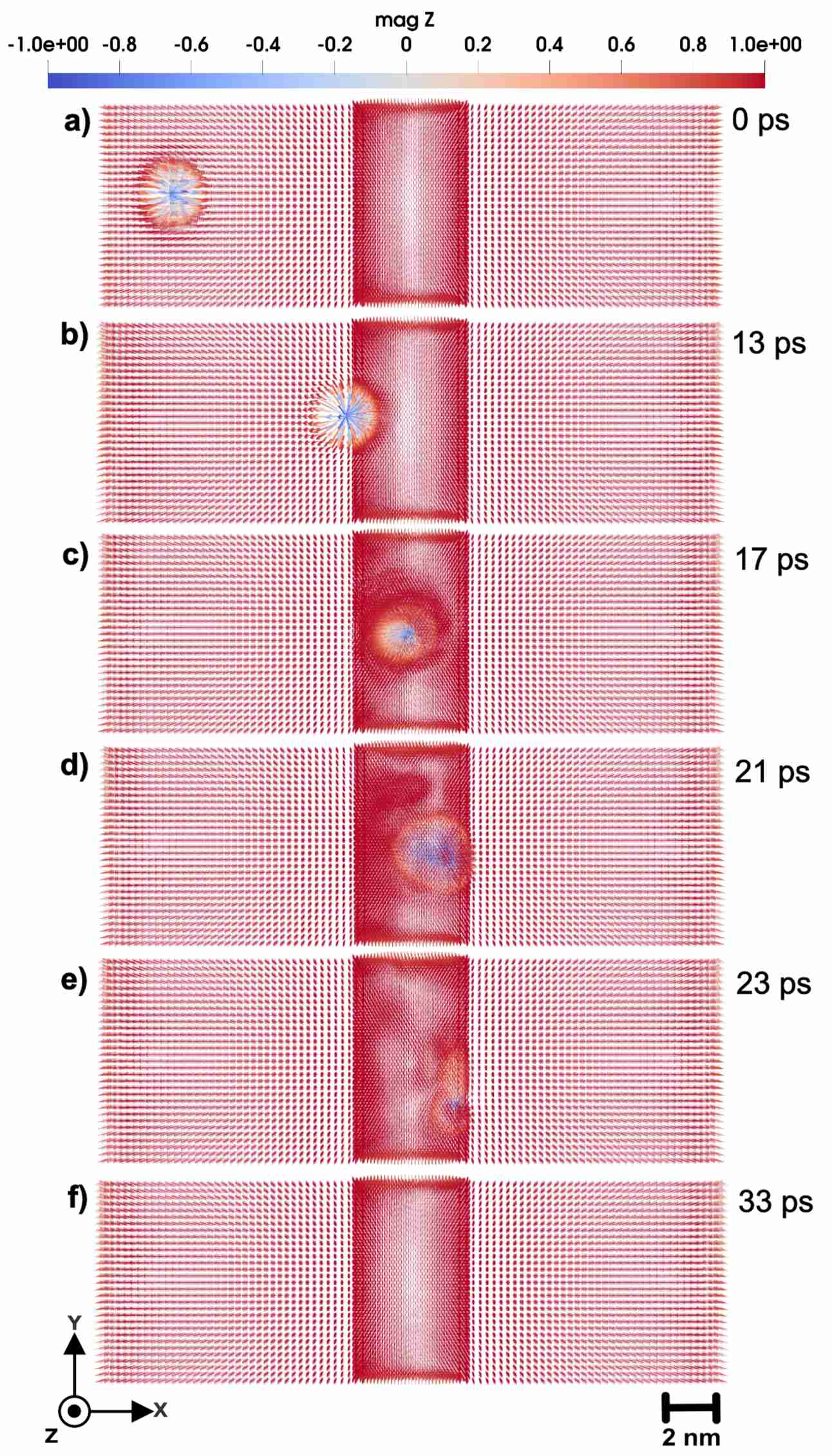}
    \caption{3D Skyrmion motion by applying STT of 25~m/s through the simulation cell, with a defect region shaped as a tetrahedron cluster with uniaxial anisotropy and hard axis along $z$-direction with strength of 0.9 mRy. The color bar is showing the $z$ component of normalized magnetization. The dynamics of the system is presented in Supplemental Material, Video 19 \cite{Supplementary}}
    \label{fig:s8}
\end{figure}

\begin{figure}[H]
    \centering
\includegraphics[width=0.44\textwidth, height=0.7\textwidth]{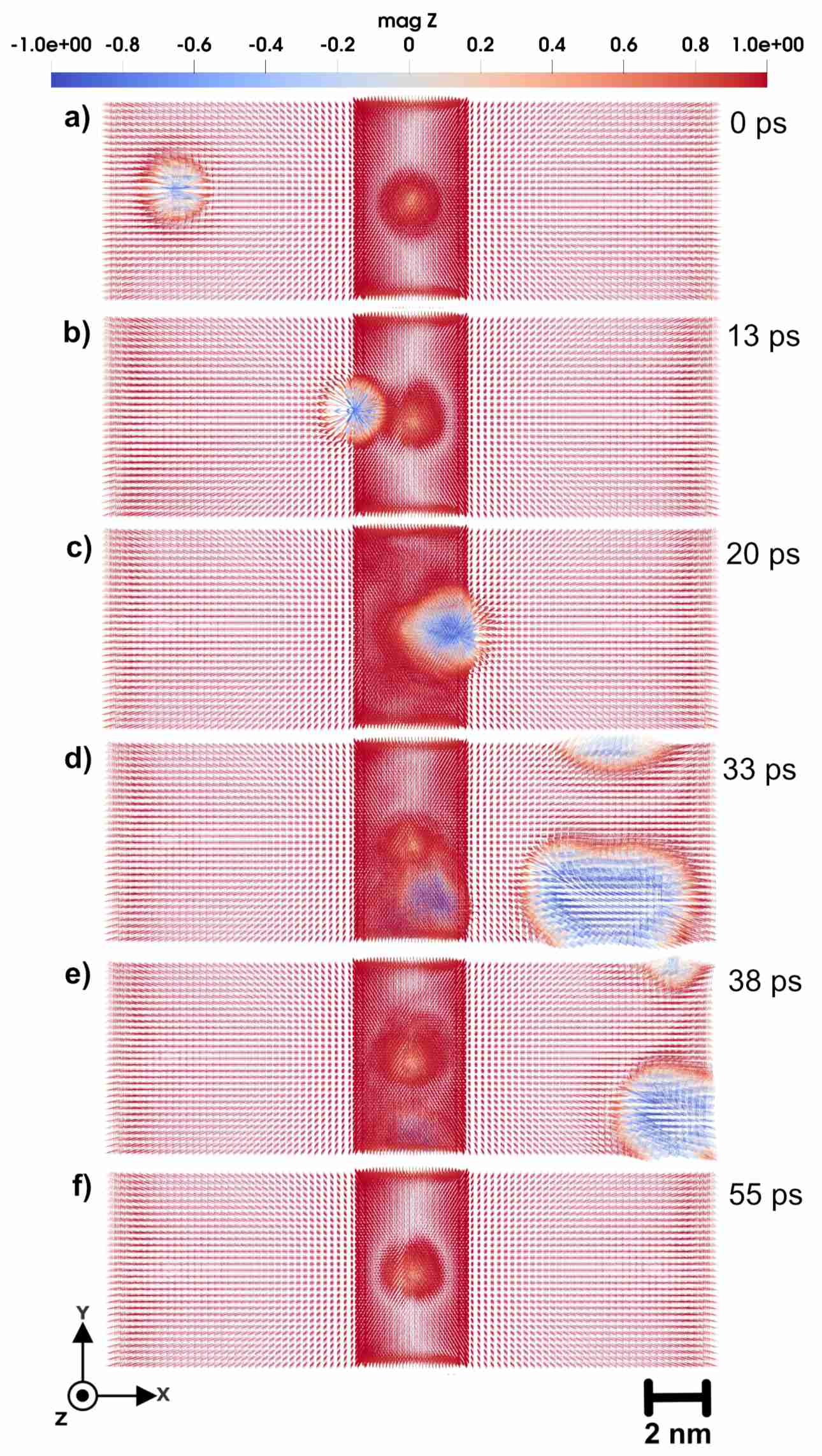}
    \caption{3D Skyrmion motion by applying STT of 25~m/s through the simulation box, with a defect region shaped as a tetrahedron cluster with uniaxial anisotropy and easy axis along $z$-direction with strength of 0.9 mRy. The color bar is showing the $z$ component of normalized magnetization. The dynamics of the system is presented in Supplemental Material, Video 20 \cite{Supplementary}}
    \label{fig:s9}
\end{figure}

%\bibliography{Supplementary/Bib_supp}

%

\end{widetext}

\end{document}